\documentclass[amsmath,amssymb,twocolumn,aps,prb,superscriptaddress]{revtex4-1}

\usepackage{mathtools}
\usepackage{mathrsfs, amsmath, wasysym}
\usepackage[mathscr]{eucal}
\usepackage{xfrac}
\usepackage{graphicx}
\usepackage{dcolumn}
\usepackage{bm}
\usepackage{color}
\usepackage{tabularx}
\usepackage{natbib}
\usepackage[colorlinks,linkcolor=blue,citecolor=blue,urlcolor=blue]{hyperref}
\usepackage[normalem]{ulem}
\usepackage[all]{hypcap}
\usepackage[dvipsnames,usenames,table]{xcolor}
\usepackage{chngcntr}

\newcommand{\nocontentsline}[3]{}
\newcommand{\tocless}[2]{\bgroup\let\addcontentsline=\nocontentsline#1{#2}\egroup}

\newcommand{\bk}{{\bf k}}

\newcommand{\bp}{{\bf p}}

\newcommand{\be}{\begin{equation}}
\newcommand{\ee}{\end{equation}}
\newcommand{\beg}{\begin{gather}}
\newcommand{\eeg}{\end{gather}}
\newcommand{\beq}{\begin{eqnarray}}
\newcommand{\eeq}{\end{eqnarray}}
\newcommand{\bea}{\begin{align}}
\newcommand{\eea}{\end{align}}
\newcommand{\beqq}{\begin{eqnarray*}}
\newcommand{\eeqq}{\end{eqnarray*}}

\newcommand{\bra}[1]{\langle #1 | }
\newcommand{\ket}[1]{ | #1 \rangle }

\newcommand{\kz}{\text{k}_\text{z}}

\newcommand{\GaAsBi}{\text{GaAs}_{0.5} \text{Bi}_{0.5}}
\newcommand{\GaAsBix}{\text{GaAs}_{1-x} \text{Bi}_{x}}
\newcommand{\GaAsBifirst}{\text{GaAs}_{0.75} \text{Bi}_{0.25}}
\newcommand{\GaAsBisecond}{\text{GaAs}_{0.25} \text{Bi}_{0.75}}
\newcommand{\GaSbBi}{\text{GaSb}_{0.5} \text{Bi}_{0.5}}
\newcommand{\InSbBi}{\text{InSb}_{0.5} \text{Bi}_{0.5}}
\newcommand{\InAsSb}{\text{InAs}_{0.5} \text{Sb}_{0.5}}
\newcommand{\InPSb}{\text{InP}_{0.5} \text{Sb}_{0.5}}

\begin{document}

\title{Ideal near-Dirac triple-point semimetal in III-V semiconductor alloys}

\author{Zhenyao Fang}
\affiliation{Department of Chemistry, University of Pennsylvania, Philadelphia, Pennsylvania 19104--6323, USA}

\author{Heng Gao}
\affiliation{Beijing National Laboratory for Condensed Matter Physics,
Institute of Physics, Chinese Academy of Sciences, Beijing 100190, China}%
\affiliation{Songshan Lake Materials Laboratory, Dongguan, Guangdong 523808, China}

\author{J\"orn W. F. Venderbos}
\affiliation{Department of Physics and Astronomy, University of Pennsylvania, Philadelphia, Pennsylvania 19104, USA}%
\affiliation{Department of Chemistry, University of Pennsylvania, Philadelphia, Pennsylvania 19104--6323, USA}

\author{Andrew M. Rappe}
\affiliation{Department of Chemistry, University of Pennsylvania, Philadelphia, Pennsylvania 19104--6323, USA}

\begin{abstract}
Despite the growing interest in topological materials, the difficulty of experimentally synthesizing and integrating them with other materials has been one of the main barriers restricting access to their unique properties. Recent advances in synthesizing metastable phases of crystalline materials can help to overcome this barrier and offer new platforms to experimentally study and manipulate band topology. Because III-V semiconductors have a wide range of functional material applications (including optoelectronic devices, light-emitting diodes, and highly efficient solar cells), and because Bi-doped III-V materials can be synthesized by ion plantation and ion-cutoff methods, we revisit the effect of bismuth substitution in metastable III-V semiconductors. Through first-principles calculation methods, we show that in wurtzite structure III-V materials, Bi substitution can lead to band inversion phenomena and induce nontrivial topological properties. Specifically, we identify that GaBi and InBi are Dirac-Weyl semimetals, characterized by the coexistence of Dirac points and Weyl points, and $\GaAsBi, \GaSbBi, \InSbBi$ are triple-point semimetals, characterized by two sets of "near Dirac" triple points on the Fermi level. These experimentally-accessible bismuth-based topological semimetals can be integrated into the large family of functional III-V materials for  experimental studies of heterostructures and future optoelectronic applications.
\end{abstract}

\maketitle

\section{Introduction \label{sec:intro}}
The interplay between topology and solid-state materials has given rise to numerous novel properties. Topological semimetals~\cite{Armitage:2018dg, Young:2012kz, Wang:2012ds, Heng:2019kd}, for instance, are characterized by high-mobility charge carriers~\cite{Borisenko:2014ed, Burkov:2016hj} (which can potentially enable high-performance photovoltaic devices, \cite{Osterhoudt:2019ju, Zhu:2017hf} solar cells, \cite{Liu:2018du} and photodiode detectors for telecommunication industry~\cite{Norton:2002, Chan:2017ef}) and exotic electromagnetic responses arising from their nontrivial topological nature \cite{Vazifeh:2013fe} (including the gyrotropic magnetic effect~\cite{Zhong:2016dja}, second-harmonic generation~\cite{McIver:2012jp, Shi:2016yw},  the chiral magnetic effect, and large magnetoresistance~\cite{Kharzeev:2014fra, Son:2013jza, Son:2012eoa}, which are useful in information storage).

In order to unlock the potential of these applications, it is necessary to find or synthesize robust and versatile material realizations of topological semimetals. These materials would ideally offer processing and manufacturing possibilities similar to hallmark III-V semiconductor materials such as GaAs or InSb. One promising strategy is to start with the class of III-V materials and search for topological semimetals within this class using compositional substitution or alloying techniques. A first step in this direction has been taken by theoretically considering the effect of bismuth substitution in zincblende III-V materials~\cite{Huang:2014hg} on the electronic structure. In the case of zincblende GaBi and InBi it was shown that Bi substitution induces a band inversion and generates a topological semimetal similar to HgTe.\cite{Bernvig:2006xw} With proper amount of Bi substitution, applying uniaxial strain then provides access to the topological insulator phase. Furthermore, in zincblende materials with CuPt type-B ordering, including $\InPSb, \InAsSb$, the spin-orbit coupling strength, which is originally linear in momentum, could be further augmented by local disorder. Because the local disorder does not average out in nano-scale supercells, the spin-orbit coupling strength is further enhanced, making these materials triple-point semimetals, which act as a bridge connecting Dirac semimetals and Weyl semimetals~\cite{Winkler:2016kk}. 

To further explore avenues for realizing topological phases in III-V materials, it is important to broaden the space of compounds and crystal structures, including metastable phases and alloys. In this regard, it is encouraging that various experimental approaches, such as ion-implantation methods and ion-exchange methods, provide the capability to synthesize several metastable wurtzite III-V nanostructures.\cite{Collino:2011er, Wood:2011ec, Wood:2012bs} Among the various III-V materials, only the nitride-based compounds naturally crystallize in wurtzite phase. In most other cases, including the aforementioned Bi-substituted compounds, the zincblende phase is more stable. Modern experimental techniques for material synthesis, however, make metastable phases a promising focus of potential applications.

Based on these motivations, in this work we revisit the effect of bismuth substitution by focusing on Bi-substituted III-V materials in the wurtzite phase. Previous studies on ternary LiGaGe-type materials, which crystallize in a stuffed wurtzite structure and the same space group as the wurtzite III-V materials, have shown that this general class of crystal structures can host various types of topological semimetals. For instance, SrHgPb \cite{Gao:2018bu} was predicted to a realize a Dirac-Weyl semimetal, exhibiting both Dirac and Weyl points, and CaAgBi \cite{Chen:2017gg} was predicted to host type-I and type-II Dirac points. Therefore, in this work we employ first-principles methods to study the topological properties of wurtzite III-V materials and show that wurtzite GaBi and InBi are Dirac-Weyl semimetals characterized by the coexistence of Dirac points and Weyl points. The Dirac points on the $\kz$ axis are induced by a band inversion and protected by $C_{6v}$ point group symmetry, and the six pairs of Weyl points are on the $\kz = 0$ plane; this coexistence pattern is similar to the case of SrHgPb.\cite{Gao:2018bu} We furthmore implement an alloying strategy and show that the alloys $\GaAsBi$, $\GaSbBi$, $\InSbBi$, when the point-group symmetry is reduced to $C_{3v}$, are triple-point semimetals, characterized by two sets of triple points on the $\kz$ axis in close proximity to the Fermi level. The splitting between the bands which form the triple points is negligible compared to the topologically nontrivial energy window of these materials, making these systems “near-Dirac semimetals”.

\section{III-V materials: crystal structures \label{sec:crystals}}

Whereas minerals such as AgI, ZnO, AlN, GaN, and InN naturally occur in a wurtzite crystal structure,\cite{Vurgaftman:2001bu} the well-known conventional semiconductors GaAs, GaSb, and InSb crystallize in the zincblende structure.\cite{Zhang:2013ea} Similarly, the Bi-based III-V materials studied here are most stable in the zincblende structure, except for InBi, which instead is most stable in the lead oxide structure~\cite{Ferhat:2006gq}. The wurtzite structure, which is closely related to the zincblende structure, is only metastable.\cite{Wang:2017kx}

\begin{figure}[ht]
\centering
\includegraphics[width=250pt]{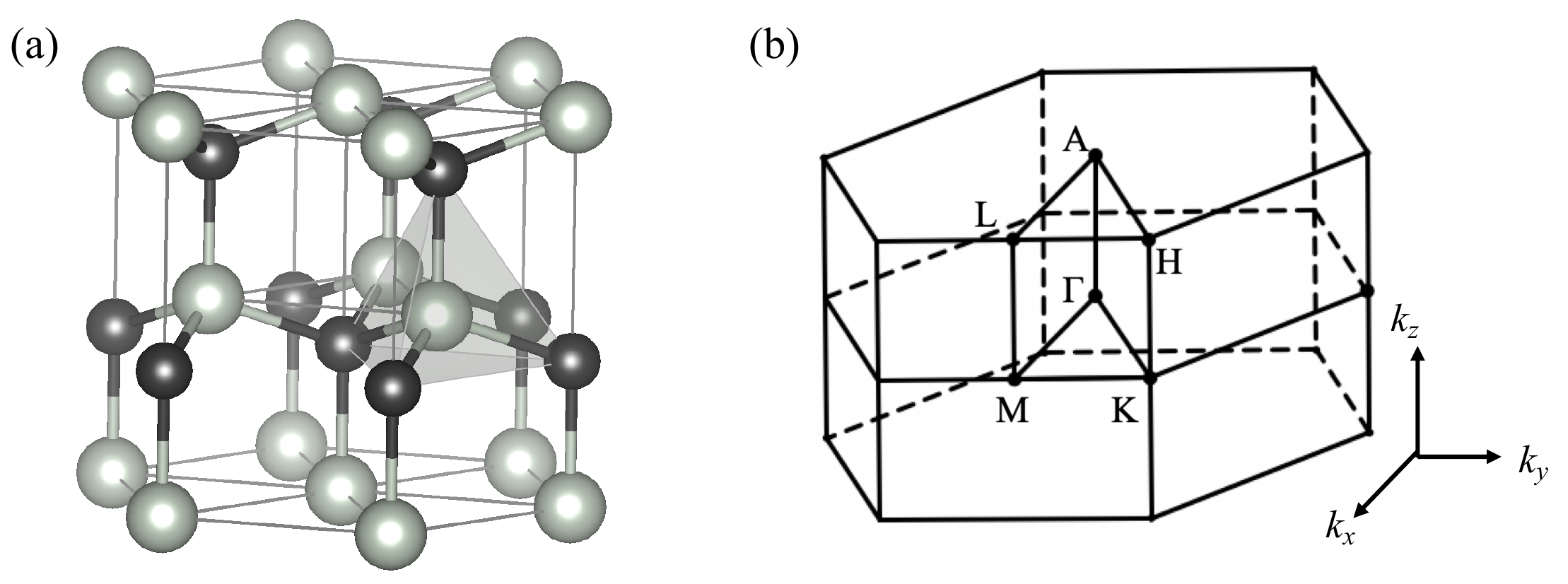}
\caption{(a) The wurtzite crystal structure, where light gray and dark gray atoms are the cations (Ga, In) and anions (As, Sb, Bi), respectively; (b) Schematic plot of the high-symmetry points in the first Brillouin zone of the wurtzite crystal structure.}
\label{fig:1}
\end{figure}

The wurtzite structure can thought of as buckled diatomic honeycomb bilayers stacked in the $c$ direction, as shown in Fig.~\ref{fig:1}. In this structure with space group $P6_3mc$ (No. 186) and associated point group $C_{6v}$, all atoms are four-fold coordinated (see Fig.~\ref{fig:1})---as is the case for the zincblende structure. The close structural and electrochemical similarity of the wurtzite and zincblende structures can be understood by comparing the view along the $(111)$ direction of the zincblende structure to the top view of wurtzite structure. The views are similar, and moreover, expose the deformation which brings the wurtzite structure into the zincblende structure. Specifically, the latter is obtained by deforming the former such that the four-fold coordinated atoms form perfect tetrahedra.

Given that the (metastable) wurtzite III-V materials are structurally similar to the zincblende materials, it is not surprising that they exhibit similar electronic properties. In particular, Bi substitution is expected to cause a band inversion in both cases, leading to topologically nontrivial electronic structure. In the case of zincblende materials this was pointed out in Ref.~\onlinecite{Huang:2014hg}; here we report the topological properties of the wurtzite compounds. 

Despite qualitative similarities, such as the presence of a band inversion, the Bi-substituted zincblende and wurtzite materials exhibit important differences in their electronic structure, which are rooted in the distinct (point group) symmetry of the two structures. In particular, the wurtzite structure is characterized by a principal six-fold screw axis (see Fig.~\ref{fig:1}), which implies that all energy bands are manifestly two-fold degenerate along the rotation axis, {\it i.e.} the $k_z$ axis in Fig.~\ref{fig:1}. As a result, an inversion of bands with different symmetry quantum numbers necessarily gives rise to a protected Dirac point crossing along the $k_z$ axis.\cite{Wang:2012ds, Armitage:2018dg} In the next section, we show that this indeed occurs in the Bi-substituted wurtzite III-V materials. 


%

\section{$\text{Bi}$-based III-V materials: Dirac-Weyl semimetal \label{sec:dirac-weyl}}

\begin{figure}[ht]
\centering
\includegraphics[width=250pt]{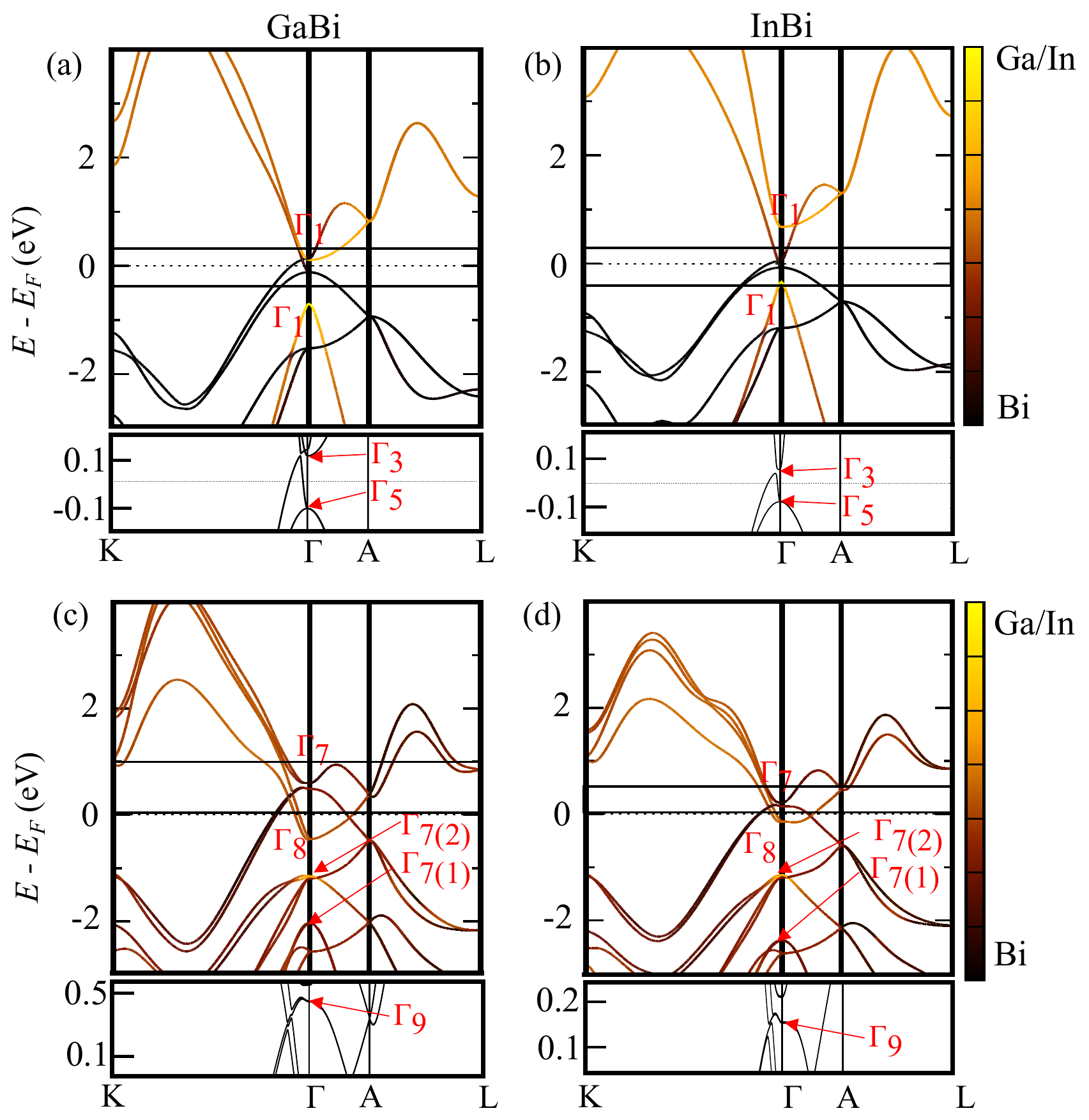}
\caption{The calculated band structures of wurtzite (a) GaBi and (b) InBi without spin-orbit coupling, and (c) GaBi and (d) InBi with spin-orbit coupling are shown along the high-symmetry lines. The inset corresponds to the boxed areas and the dashed line denotes the Fermi level.}
\label{fig:2}
\end{figure}

To obtain the electronic structure of the Bi-substituted III-V materials in the wurtzite structure, we employ first-principles density-functional theory (DFT) calculations. Details of the computational methodology are presented in Appendix~\ref{app:methods}. The energy band structures of GaBi and InBi (shown in Fig.~\ref{fig:2}), both with and without spin-orbit coupling, demonstrate the effect of Bi-substitution: a band inversion occurs involving $\Gamma_8$ and $\Gamma_9$ bands at the zone center. The $\Gamma_9$ band consists of $j_z=\pm \frac32$ $p$-states, whereas the $\Gamma_8$ band consists of $j_z=\pm \frac12$ $s$-states. The key consequence of the inverted band ordering in GaBi and InBi is the presence of symmetry-protected energy band crossings on the $\Gamma-A$ high symmetry line, located at $\bk_D = (0, 0, \pm 0.287) \, \text{\AA}^{-1}$ and $\bk_D= (0, 0, \pm 0.201) \, \text{\AA}^{-1}$, and with energies $E - E_{F} = -0.07 \, \text{eV}$ and $E - E_{F} = -0.05 \, \text{eV}$, respectively. These band crossings realize band inversion-induced Dirac points of the kind first proposed for, and observed in, $\text{Na}_3 \text{Bi}$.\cite{Wang:2012ds}


As mentioned in the previous section, the double degeneracy of all bands along the $\Gamma - A$ line can be derived from a two-fold screw rotation symmetry, which is contained by a six-fold screw axis. The Dirac point crossing is protected by the different symmetry properties of the $\Gamma_8$ and $\Gamma_9$ bands. This distinction critically relies on the six-fold (screw) rotation ({\it i.e.} point group $C_{6v}$); breaking rotation symmetry allows the bands to couple and hybridize, generating a Dirac mass and giving rise to a topological insulator phase.


\begin{figure}[ht]
\centering
\includegraphics[width=250pt]{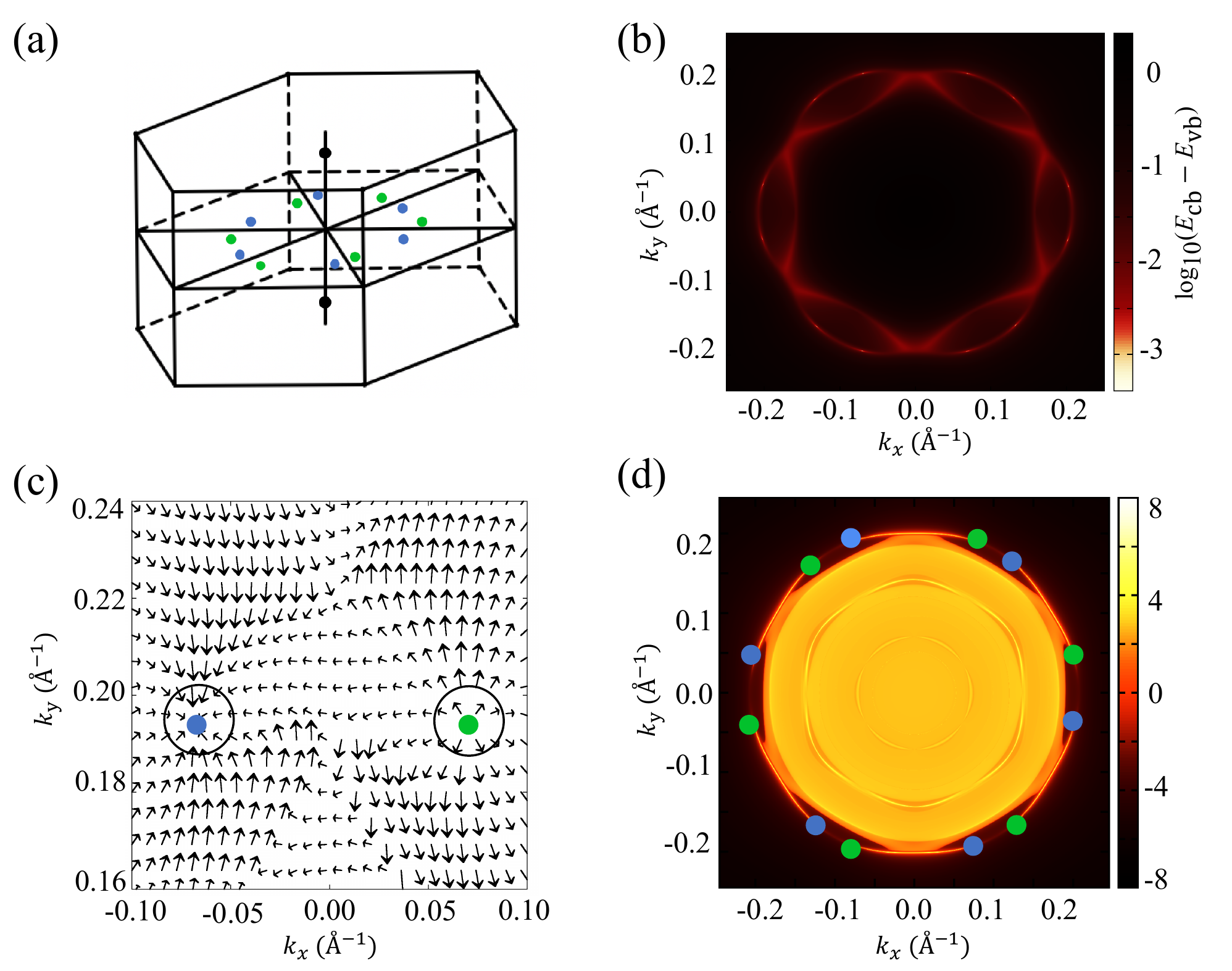}
\caption{(a) Schematic illustration of the distribution of Dirac points (black dots), Weyl points with chirality +1 (green dots) and chirality -1 (blue dots) in Dirac-Weyl semimetal GaBi. (b) The logarithmic plot of the energy difference between conduction band and valence band in the $k_z = 0$ plane. (c) Calculated Berry curvature of GaBi around two Weyl points on the $k_z = 0$ plane; the arrow denotes the in-plane projection of Berry curvature. (d) Calculated surface Fermi arcs of GaBi on the $(001)$ surface.}
\label{fig:3}
\end{figure}

In addition to the Dirac points stabilized by (screw) rotation symmetry, we also find Weyl points in the $k_z = 0$ plane, perpendicular to the screw rotation axis. The Weyl points also originate from the band inversion, as evidenced by Figs.~\ref{fig:2}(c) and (d), but do not occur on high-symmetry lines. In GaBi, for instance, one pair of the Weyl points is located at $\bk_{W_{\pm}} = (\pm 0.069, 0.194, 0) \, \text{\AA}^{-1}$, with the energy of $E - E_F = 0.156 \, \text{eV}$; the location of the remaining pairs can be obtained by rotation symmetry. To visualize the distribution pattern of Weyl points, we provide the logarithmic plot of the energy difference between the conduction band and valence band, as shown in Fig.~\ref{fig:3}(b), where the twelve light points correspond to the Weyl points. Note that the local stability of Weyl points in the $k_z = 0$ plane follows from a combined symmetry involving time-reversal and two-fold rotation.\cite{Soluyanov:2015cn} The coexistence pattern of Dirac and Weyl points is summarized in Fig.~\ref{fig:3}(a), showing both Dirac points on the $k_z$ axis (black dots) and Weyl points on the $k_z = 0$ plane (green and blue dots). Such coexistence was first predicted to occur in a class of hexagonal $ABC$ materials with the same space group in Ref.~\onlinecite{Gao:2018bu}, showcasing SrHgPb as an example. Dirac-Weyl semimetals of this kind raise the prospect of studying the interplay between Dirac and Weyl electrons, in particular considering the large distance between Weyl points in momentum space. Here we show that Bi-substituted III-V materials in the wurtzite structure are candidate materials for realizing Dirac-Weyl semimetals. 



To robustly prove the nontrivial topology of the Weyl points we determine the chiral charge by calculating the integral of Berry curvature over a sphere enclosing each Weyl point, with a radius of $0.005 \, \text{\AA}$. Specifically, we calculate $C = \frac{1}{2 \pi i} \oint d \mathbf{S} \cdot \mathbf{B} (\mathbf{k})$ with the Berry curvature $\mathbf{B} (\mathbf{k})$ defined as $\mathbf{B} (\mathbf{k}) = \nabla_{\mathbf{k}} \times \sum_{\text{occ.}} \bra{u_n (\mathbf{k})} \nabla_{\mathbf{k}} \ket{u_n (\mathbf{k})}$, where the sum is over all occupied bands labeled by $n$. The chirality of each Weyl point is shown in Fig.~\ref{fig:3}(a), where the green dots indicate chirality $C=+1$ and blue dots $C=-1$. Since Weyl points with chirality $+1$ and $-1$ are the sources and sinks of Berry curvature, respectively, we plot the Berry curvature $\mathbf{B} (\mathbf{k})$ as function of $(k_x,k_y)$ in the $k_z = 0$ plane, shown in Fig.~\ref{fig:3}(c), as a graphical representation of the nontrivial topology. The Weyl points are labeled by green and blue dots and are shown to indeed correspond to sources or sinks.

An important consequence of nontrivial bulk topology is the appearance of surface states. In the case of Weyl electrons, Fermi arc surface states must connect the projections of the Weyl points onto the surface Brillouin zone. In the present case, where the Weyl points are located in the $k_z=0$ plane, Fermi arc surface states are expected on the $(001)$ surface. To verify this, we calculate the surface spectral function at the Weyl point energy $E - E_F = 0.156 \, \text{eV}$ and show the result in Fig.~\ref{fig:3}(d). The central hexagon corresponds to the bulk states because the energy is not chosen to be the Fermi energy $E_F$, and the six lines around that hexagon, which connects Weyl points of opposite chirality, correspond to the surface Fermi arc states, which, in the case of Weyl points, are protected by topology and are robust against weak external perturbations. These topological Fermi arcs could be detected by angle-resolved photo-emission spectroscopy as the signal of the topological nature of these materials.~\cite{Lv:2015yn, Jiang:2017ba}


\section{$\text{Bi}$-based III-V alloys: triple-point semimetals \label{sec:triple_point}}

\begin{figure}[h]
\centering
\includegraphics[width=250pt]{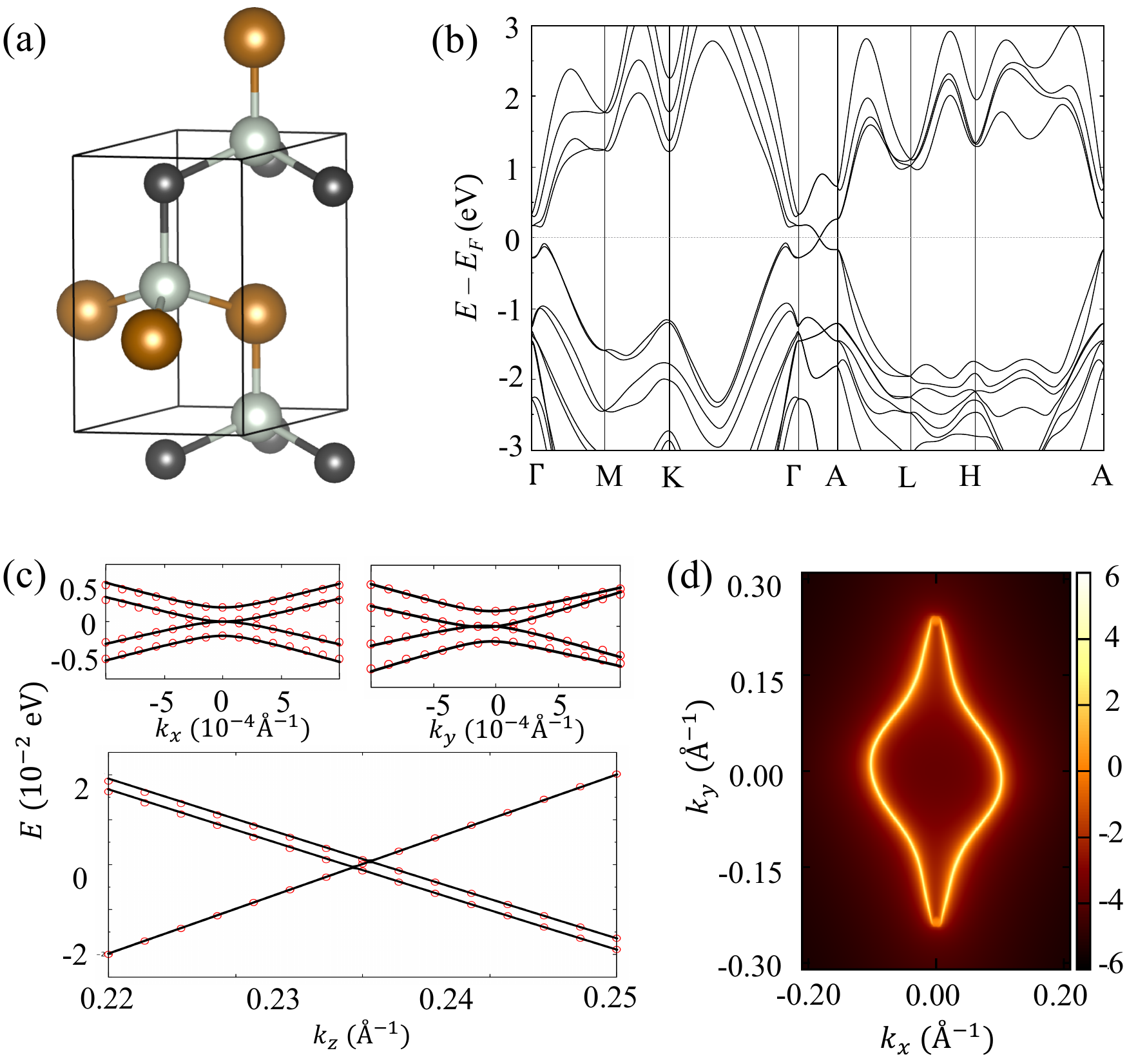}
\caption{(a) The crystal structure of the unit cell of alloy $\GaAsBi$, where light gray, dark gray, and brown atoms denote Ga, As, and Bi, respectively. (b) The calculated band structure of $\GaAsBi$, labeled by the irreducible representations of $C_{3v}$ point group, is shown along the high-symmetry line. The dashed line denotes the Fermi level. (c) The fit of the $\bk \cdot \bp$ Hamiltonian (red circles) with DFT results (black lines) in the vicinity of the triple points. (d) The calculated surface Fermi arcs of the $(100)$ plane of $\GaAsBi$.}
\label{fig:4}
\end{figure}

The presence of both Weyl and Dirac points in Bi-substituted GaAs and InSb is encouraging, yet for application purposes it is desirable to have topological band crossings right at the Fermi energy. As shown by Fig.~\ref{fig:2}, the Dirac points are below the Fermi energy and associated with an electron pocket, whereas the Weyl points are above the Fermi energy and associated with a hole pocket. Aiming to remedy this, we employ an alloying strategy inspired by similar work on zincblende III-V materials~\cite{Huang:2014hg}. By alloying normal insulators GaAs, GaSb, and InSb with Dirac-Weyl semimetals GaBi and InBi the series of alloys $\GaAsBi$, $\GaSbBi$, and $\InSbBi$ can be obtained; the crystal structures of $\GaAsBi$, which is taken as an example, is shown in Figure~\ref{fig:4}(a).

The main effect of alloying is to tune the strength of spin-orbit coupling, which directly affects the band inversion. A material which is still inverted but closer to the band inversion transition is likely to exhibit protected Dirac band crossings at the Fermi energy, without additional Fermi surface. We find that this is indeed the case for $\GaAsBi$, as shown in Fig.~\ref{fig:4}(b): the crossing of energy bands on the $\Gamma - A$ line is now at the Fermi energy. Alloying, however, also reduces the point group symmetry; in case of $\GaAsBi$ and similar alloys the symmetry is reduced from $C_{6v}$ to $C_{3v}$. This has implications for the degeneracy of energy bands, in particular on the rotation axis, and affects the protection of Dirac points. When rotation symmetry is reduced to three-fold, the $j_z = \frac32$ states are no longer degenerate on the $k_z$ axis but instead are split (except for the time-reversal invariant points). As the result, the Dirac points are split into triple points, giving rise to a triple-point topological semimetal.\cite{Zhu:2016cb} In the case of $\GaAsBi$ this is clearly shown in the bottom panel of Fig.~\ref{fig:4}(c).


The electronic states in the vicinity of the triple points can be well-described by the low-energy $\bk \cdot \bp$ Hamiltonian derived in Ref.~\onlinecite{Winkler:2016kk}. Adapted to the present case, the Hamiltonian is given by

\footnotesize
\begin{equation}
    H(\bk)=
    \begin{pmatrix}
    E_0 + A k_z & 0 & D k_x & D k_y \\
    0 & -E_0 + A k_z & F^* k_y & -F^* k_x \\
    D^* k_x & F k_y & B k_z + C k_y & C k_x \\
    D^* k_y & -F k_x & C k_x & B k_z - C k_y 
    \end{pmatrix} \label{eq:H}
\end{equation}
\normalsize

\noindent where here $k_x$ and $k_y$ are interchanged with respect to Ref.~\onlinecite{Winkler:2016kk}, which is due to a different choice of coordinate system. We fit the parameters of \eqref{eq:H} to the first-principles band structure calculations and show the result in Fig.~\ref{fig:4}(c), which demonstrates that \eqref{eq:H} provides a good description of the low-energy electronic structure. The values of the fit parameters are listed in Table~\ref{tab:parameters}. 


\begin{table}[t]
\centering 
\begin{ruledtabular}
\begin{tabular}{ccc}
$A$ (eV \AA) & $B$ (eV \AA) & $C$ (eV \AA) \\ \hline
-2.000 & 2.286 & 0.754 \\ 
$D$ (eV \AA) & $E_0$ (meV) & $F$ (eV \AA) \\  \hline
2.413 & 1.242 & -2.485 
\end{tabular}
\end{ruledtabular}
\caption{The fitted parameters of the $\bk \cdot \bp$ Hamiltonian.}
\label{tab:parameters} 
\end{table}

As suggested by Fig.~\ref{fig:4}(b), Fig.~\ref{fig:4}(c) shows that the splitting of bands on the $k_z$ axis as a result of symmetry lowering is small in magnitude. In fact, we find that the splitting is $\Delta E \approx 3 \, \text{meV}$. Such small splitting and close proximity of the triple points leads to the expectation that semimetals of this kind behave as Dirac semimetals, motivating the term “near-Dirac semimetal”. Since the near-Dirac point is at the Fermi energy, with no additional electron or hole pockets, such near-Dirac semimetals provide a promising venue for observing signatures of and studying the properties of Dirac semimetals. For instance, it is expected that the Fermi arc surface states can be clearly resolved. To verify this, we calculate the surface spectral function of the $\GaAsBi$ $(100)$ alloy surface and show the result in Fig.~\ref{fig:4}(d), which corroborates that the spectral function closely resembles that of a proper band-inverted Dirac semimetal.\cite{Armitage:2018dg}


It is worth noting that for substitution proportions $x = 0.25, 0.75$ in $\GaAsBix$ a supercell with $C_{6v}$ symmetry can be constructed, as shown in Appendix~\ref{app:C6v}. For $x=0.5$ that is not the case, reducing the symmetry to $C_{3v}$. In the case of the former, four-fold degenerate Dirac points can be strictly protected by symmetry, and we therefore examine the band structures of $\GaAsBifirst$ and $\GaAsBisecond$ for an ideal Dirac point crossing. We find that Dirac points exist in both materials; details of them could be found in Appendix~\ref{app:C6v}. Combined with the band structures of GaAs, $\GaAsBi$, and GaBi, these band structures could further demonstrate the effect of bismuth alloying.

\begin{figure}[htb]
\centering
\includegraphics[width=250pt]{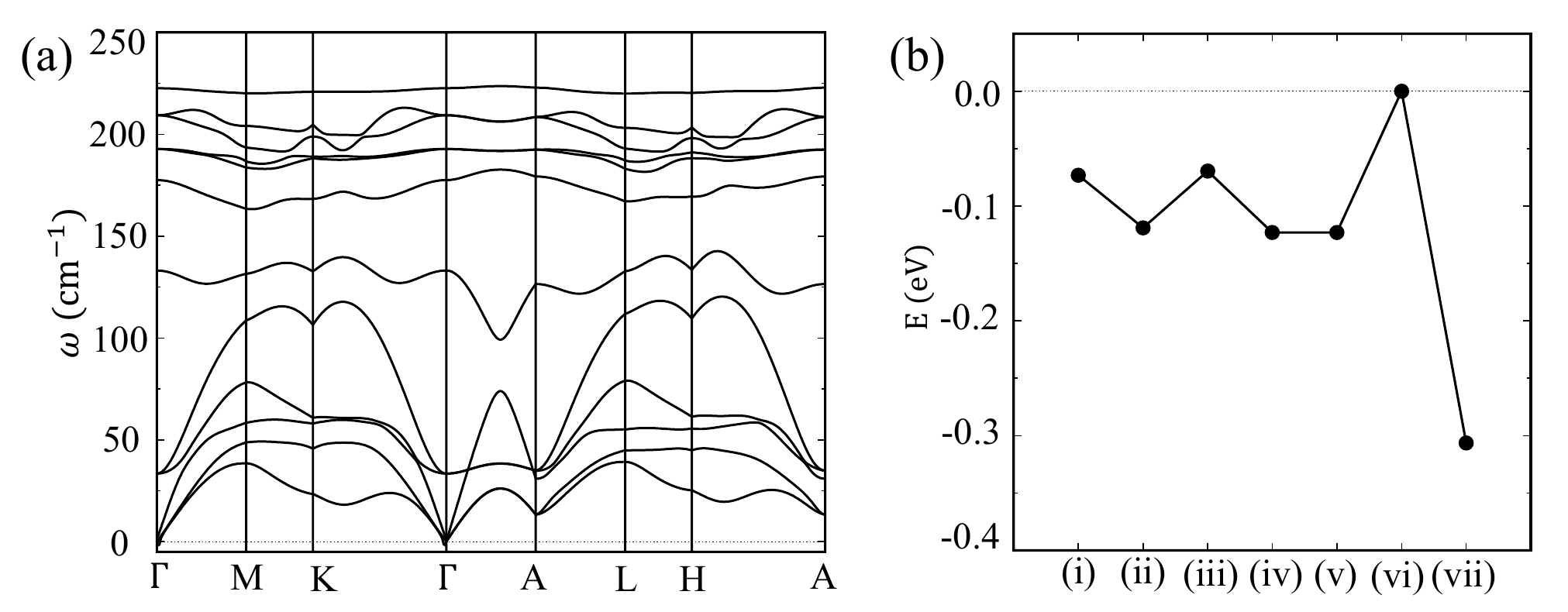}
\caption{(a) The calculated phonon dispersion relations of $\GaAsBi$ is shown along the high-symmetry lines of the Brillouin zone of the hexagonal lattice. (b) The relative energies of different alloying orders of the $2 \times 2 \times 1$ supercell $\GaAsBi$. The lowest-energy ordering (vii) is the triple-point near-Dirac semimetal.}
\label{fig:5}
\end{figure}

To show the thermodynamic stability of $\GaAsBi$, we first calculate its phonon dispersion relation. The calculated phonon modes at $\Gamma$ point have a small negative mode $\omega = -17.18 \, \text{cm}^{-1}$ without using acoustic sum rules. After applying the sum rules, the negative mode is eliminated and the phonon dispersion relation is everywhere positive, as shown in Figure~\ref{fig:5}(a), showing that $\GaAsBi$ is a metastable crystalline material. Furthermore, we consider the energies of all alloy orderings in a $2 \times 2 \times 1$ $\GaAsBi$ supercell; the seven orderings are listed in Appendix~\ref{app:alloys}, among which the ordering (vii) is the desired triple-point near-Dirac semimetal with $C_{3v}$ group symmetry. The energies of these alloying orders are shown in Figure~\ref{fig:5}(b). From Figure~\ref{fig:5}(b), it is shown that the energy of ordering (vii) is lower than the other ordering by at least $0.18 \, \text{eV}$ per unit cell, which suggests that it is energetically favorable to obtain the proposed triple-point semimetal compared with other disordered crystalline phases which do not host nontrivial topology. In conclusion, these results could show that the triple-point near-Dirac semimetal structure of $\GaAsBi$ is more stable than other alloy ordering and could be realized in experiments.

\section{Conclusion}
In this work, we revisited the effect of bismuth substitution to lead to nontrivial topological properties in metastable wurtizte III-V materials. Based on first-principles calculations, full substitution of bismuth would lead to Dirac-Weyl semimetals GaBi and InBi, characterized by the coexistence of Dirac points and Weyl points in the Brillouin zone. By alloying them with normal insulators, the reduction of the spin-orbit coupling strength could remove the additional hole pockets near the Brillouin zone center, and the change of crystal symmetry splits the Dirac points into a set of triple points, making the thermodynamically metastable alloys $\GaAsBi$, $\GaSbBi$, and $\InSbBi$ triple point semimetals. The integration of topological properties with conventional wurtzite III-V materials, along with the advanced experimental approaches to synthesizing them, could enrich the family of accessible topological materials and provide new platforms for people to experimentally study the various properties arising from topology.



\section{Acknowledgements}
We would like to thank Prof. Ruixiang Fei for helpful disucssions and advice concering the calculations. Z. F. and A. M. R. acknowledge support from the National Science Foundation, under grant number DMR-1808202. H.G. acknowledges support from National Postdoctoral Program for Innovative Talents under Grant No. BX20190361. J. W. F. V. acknowledges support from the National Science Foundation, under grant number DMR-1720530.

\appendix
\counterwithin{figure}{section}
\counterwithin{table}{section}
\section{Methodology \label{app:methods}}
In this work, calculations were performed based on density functional theory implemented in the software of Quantum Espresso \cite{Giannozzi:2009hx}. For structural relaxation, since the experimental values of the lattice constants of wurtzite III-V materials are not readily available while those of zincblende materials are,\cite{Vurgaftman:2001bu, Tixier:2003jc, Shalindar:2016km} the initial lattice constants are obtained by the relations $a_{\text{WZ}} = \frac{1}{\sqrt{2}} a_{\text{ZB}}$ and $c_{\text{WZ}} = \sqrt{\frac{3}{8}} a_{\text{WZ}}$. Then the lattice constants of wurtzite materials were obtained through structural relaxation with the strongly constrained and appropriately normed (SCAN) semilocal meta-GGA functional.\cite{Sun:2015ef} 

The electronic band structures were calculated with the Heyd-Scuseria-Ernzerhof (HSE) hybrid functional \cite{Heyd:2003eg}, since GGA or LDA functionals are known to underestimate band gaps of semiconductors.\cite{Cohen:2011fm} In the band structure calculations, a $6 \times 6 \times 6$ $ \mathbf{k}$-point grid and an energy cutoff of $50 \, \text{Ry}$ could give converged results and are thus used. Furthermore, phonon dispersion relation calculations are performed to show the thermodynamic stability of these wurtzite materials; they were calculated with a $4 \times 4 \times 2$ $ \mathbf{q}$-point grid and an energy cutoff of $100 \, \text{Ry}$. 

In order to study the topological properties of the proposed materials, the Hamiltonian of each material was constructed based on maximally-localized Wannier functions, obtained in the software of Wannier90,\cite{Mostofi:2014hw} with the projections of the bands around the Fermi level chosen to be the s-orbital states of the cations and the p-orbital states of anions, as used in previous work.\cite{Gmitra:2016gf,Kim:2009dj} The validity of this choice of projection is shown by comparing the band structures obtained in Quantum Espresso and Wannier90. Furthermore, the computational software of Wanniertools\cite{Wu:2018en} is used to obtain the topological properties, such as the surface Fermi arcs, the chirality of Weyl points, and Berry curvature, of these materials, based on the Hamiltonian obtained from the software of Wannier90. 

\section{Absence of bismuth substitution - normal insulators \label{app:no-Bi}}
\begin{figure}[hb]
\centering
\includegraphics[width=215pt]{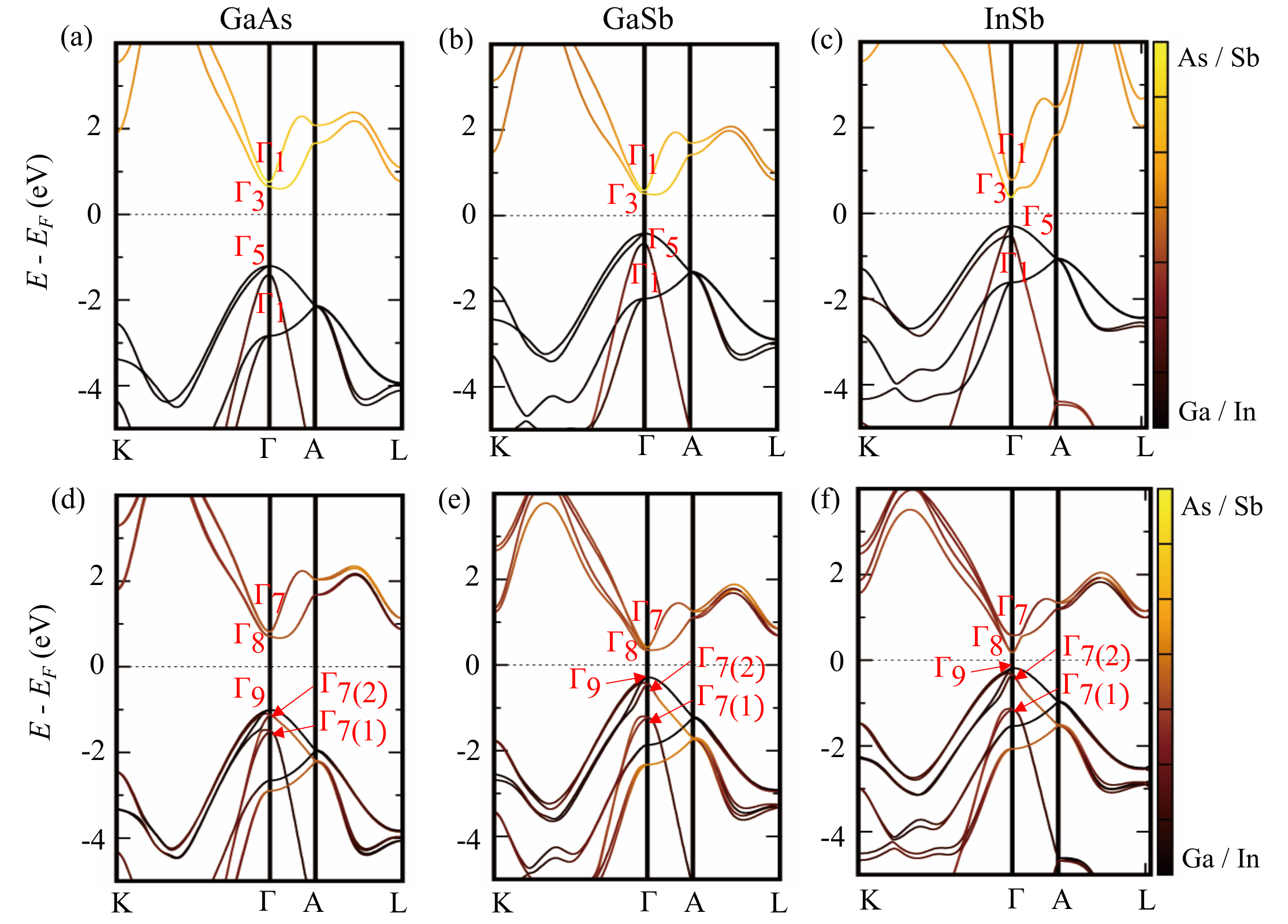}
\caption{The calculated band structures of wurtzite (a) GaAs, (b) GaSb, and (c) InSb without spin-orbit coupling, and (d) GaAs, (e) GaSb, and (f) InSb with spin-orbit coupling are shown along the high-symmetry lines. The dashed line denotes the Fermi level.}
\label{figa1}
\end{figure}

Without bismuth substitution, the wurtzite III-V materials, GaAs, GaSb, and InSb, are normal insulators, as shown in the band structures in Figure~\ref{figa1}. To further analyze these band structures, we define the spin-orbit coupling strength of the anions in these materials as the energy difference between $p_{1/2}$- and $p_{3/2}$-states. Under $C_{6v}$ group symmetry, $p_{3/2}$-states will split into $\Gamma_7$ and $\Gamma_9$ bands, formed by $|\frac{3}{2}, \pm \frac{1}{2} \rangle$ and $|\frac{3}{2}, \pm \frac{3}{2} \rangle$ states respectively. To distinguish the $\Gamma_7$ band originating from $p_{1/2}$ and $p_{3/2}$ states, we label them by $\Gamma_{7(1)}$ and $\Gamma_{7(2)}$. Therefore, the spin-orbit coupling strength is expressed as $E_{\text{soc}} = \frac{1}{2} (E_{\Gamma_9} + E_{\Gamma_{7(2)}}) - E_{\Gamma_{7(1)}}$. The energy band gap $E_{\text{gap}}$ and $E_{\text{soc}}$ of GaAs, GaSb, and InSb can be found in Table~\ref{tab:band_parameters}. These values are in accordance with those being reported before,\cite{Gmitra:2016gf} thus verifying the validity of our calculations.

For GaBi and InBi, however, the band gap cannot be defined by the energy difference between $\Gamma_8$ and $\Gamma_9$ bands. Therefore, we could identify the energy difference between $\Gamma_9$ and $\Gamma_8$ bands as the band inversion strength. Instead of introducing $E_{\text{bis}} = E_{\Gamma_9} - E_{\Gamma_8}$, as what was done in Ref.~\onlinecite{Huang:2014hg}, we will continue to use the definition of $E_{\text{gap}}$ since both quantities only involve the energies of $\Gamma_9$ and $\Gamma_8$ bands. Therefore, negative band gap indicates that $\Gamma_9$ and $\Gamma_8$ bands are inverted in our case. These values could also be found in Table~\ref{tab:band_parameters}.

\begin{table}[t]
\centering 
\begin{ruledtabular}
\begin{tabular}{cccccc}
 & GaAs & GaSb & InSb & GaBi & InBi \\  \hline
$a_{\text{WZ}} \,$(\AA) & 3.866 & 4.428 & 4.498 & 4.397 & 4.657 \\
$c_{\text{WZ}} \,$(\AA) & 6.386 & 7.028 & 7.433 & 7.136 & 7.543 \\
$E_{\text{gap}} \, (\text{eV})$ & 1.72 & 0.64 & 0.36 & -0.91 & -0.68 \\
$E_{\text{soc}} \, (\text{eV})$ & 0.50 & 0.94 & 0.97 & 2.54 & 2.54 \\
\end{tabular}
\end{ruledtabular}
\caption{Calculated lattice constants $a_{\text{WZ}}$ and $c_{\text{WZ}}$, energy band gap $E_{\text{gap}}$, spin-orbit coupling strength $E_{\text{soc}}$ of wurtzite GaAs, GaSb, InSb, GaBi and InBi}
\label{tab:band_parameters} 
\end{table}

\section{The different alloy ordering of $2 \times 2 \times 1$ supercell of wurtzite $\GaAsBi$ \label{app:alloys}}
\begin{figure}[ht]
\centering
\includegraphics[width=250pt]{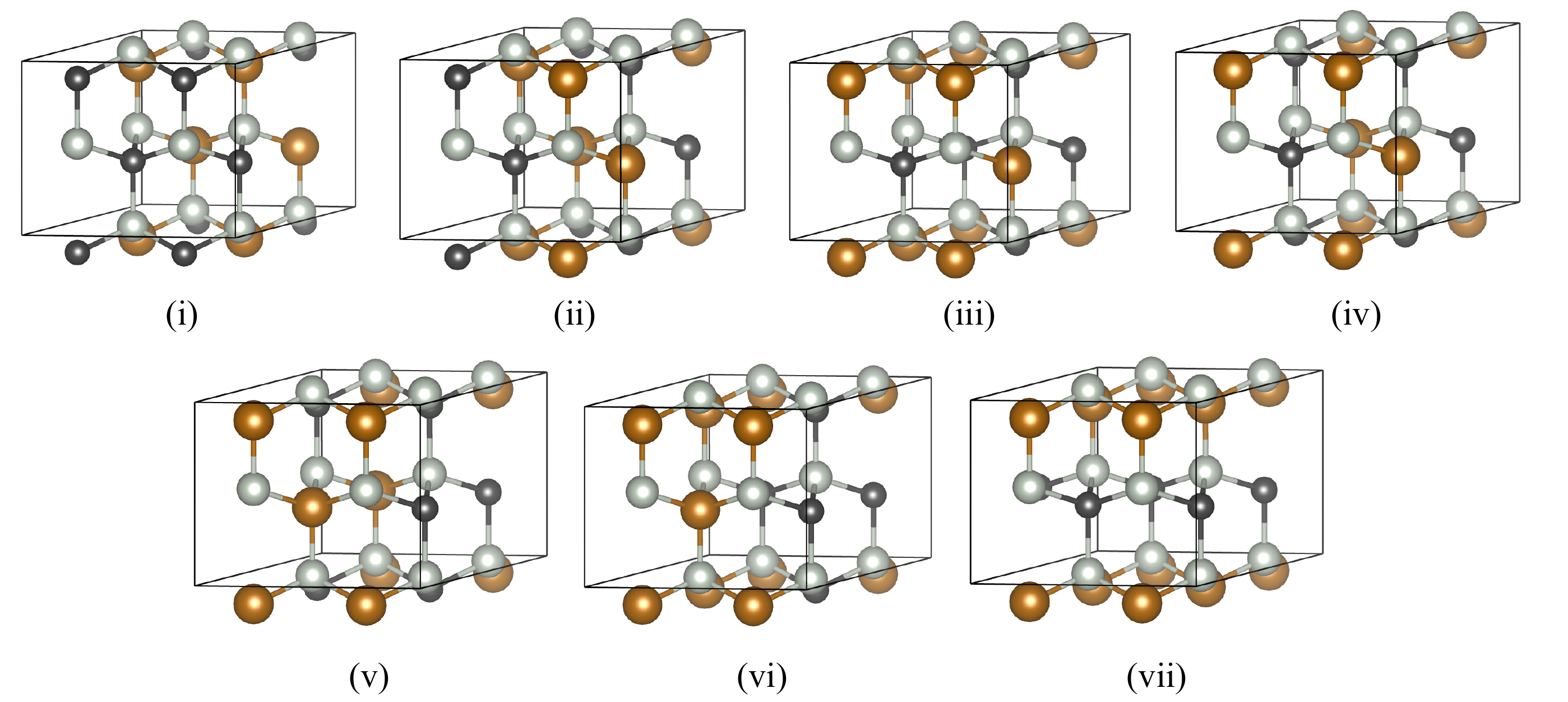}
\caption{The different alloy ordering of $2 \times 2 \times 1$ supercell $\GaAsBi$. The ordering (vii) is the desired triple-point near-Dirac semimetal.}
\label{figa2}
\end{figure}

In this appendix, the crystal structures of different alloying orders of $\GaAsBi$ in a $2 \times 2 \times 1$ supercell are listed. There are in total seven inequivalent crystal structures, shown in Figure~\ref{figa2}, with light gray, dark gray, and brown atoms representing Ga, As, and Bi, respectively.

\section{The crystal structure and band structures of $\GaAsBifirst$ and $\GaAsBisecond$ \label{app:C6v}}

\begin{figure}[ht]
\centering
\includegraphics[width=200pt]{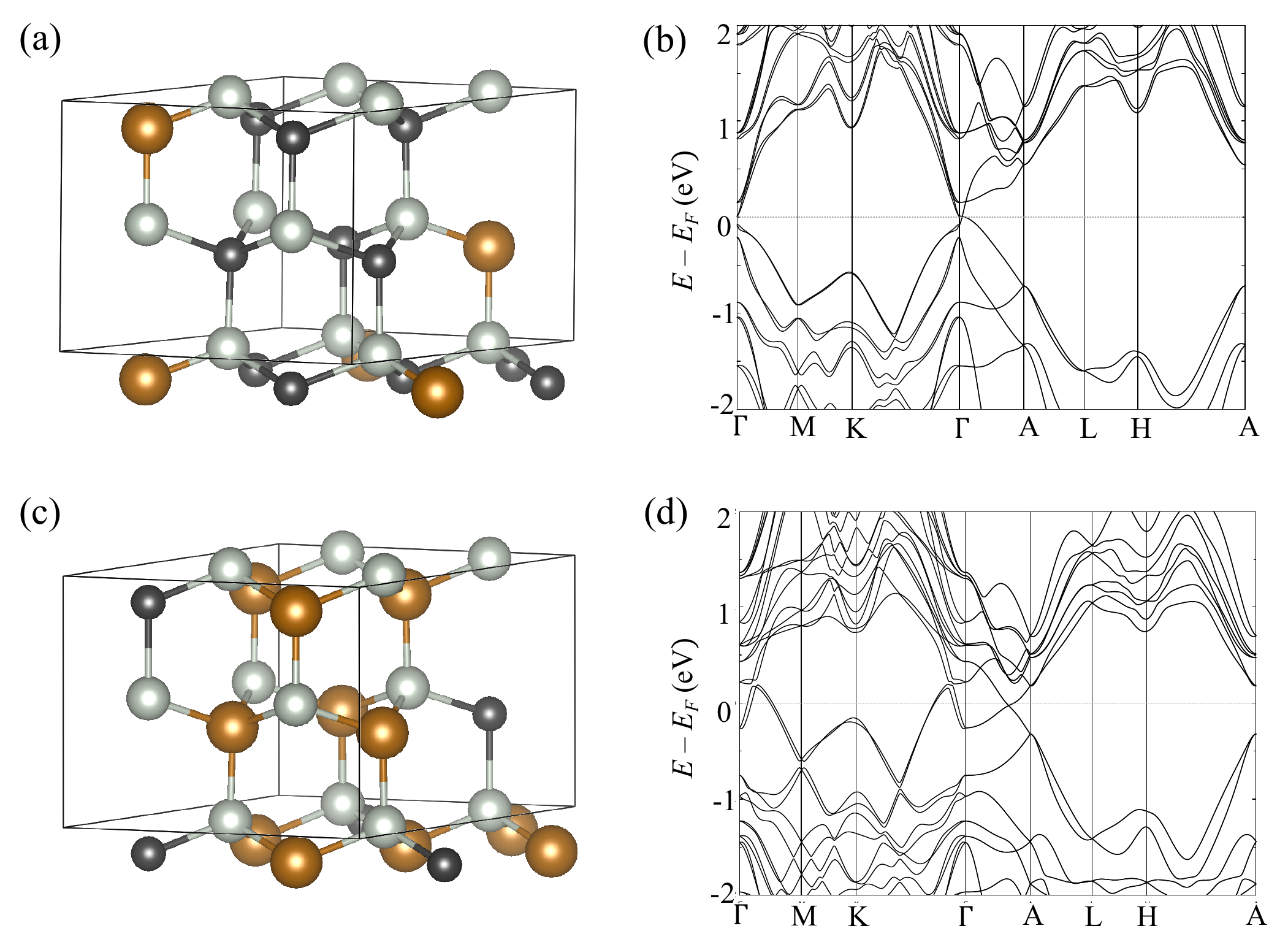}
\caption{The crystal structures and calculated band structures of $\GaAsBifirst$ (a, b) and $\GaAsBisecond$ (c, d). Here light gray, dark gray, and brown atoms represent Ga, As, and Bi, respectively. The dashed line denotes the Fermi level.}
\label{figa3}
\end{figure}

For $\GaAsBifirst$ and $\GaAsBisecond$, we can construct $2 \times 2 \times 1$ supercell with $C_{6v}$ crystal symmetry, as is shown in Figure~\ref{figa3}, while for $\GaAsBi$ this is not possible. From the calculated band structures, we could further demonstrate the effect of bismuth substitution in III-V materials - to increase spin-orbit coupling and to induce band inversion. This can be seen from the positions of the Dirac point or from the band inversion strength defined in Appendix~\ref{app:no-Bi}. As is shown in Table~\ref{tab:dirac_point}, when substituting bismuth into GaAs, which has no band inversion and Dirac points, the spin-orbit coupling strength gradually increases, thus the band inversion strength should increase and the Dirac points move from $\Gamma$ point to $A$ point.

Although Dirac points exist in both $\GaAsBifirst$ and $\GaAsBisecond$, the bands forming the Dirac point are different. In $\GaAsBifirst$, $\Gamma_9$ band becomes higher than the Fermi energy due to its stronger spin-orbit coupling strength than $\GaAsBi$, but its change is not sufficient enough to induce a band inversion between $\Gamma_9$ and $\Gamma_8$ bands, so the Dirac point is formed by $\Gamma_9$ and $\Gamma_7$ bands. And in $\GaAsBisecond$, the Dirac point does come from band inversion between $\Gamma_9$ and $\Gamma_8$ bands, which is also the case for GaBi. Similarly, it is not on the Fermi level because of the additional electron pockets on the Fermi level.

\begin{table}[ht]
\centering 
\begin{ruledtabular}
\begin{tabular}{cccccc}
$x$ & 0.00 & 0.25 & 0.50 & 0.75 & 1.00  \\  \hline
$E_{\text{soc}} \, (\text{eV}$) & 0.50 & 0.94 & 1.26 & 1.57 & 2.54 \\
$E_{\text{gap}} \, (\text{eV}$) & 1.72 & 0.14 & -0.45 & -0.63 & -0.91\\
$(k_z)_D \,$(\AA) & -- & 0.02 & 0.237 & 0.269 & 0.287 \\
\end{tabular}
\end{ruledtabular}
\caption{The spin-orbit coupling strength, the energy band gap, and the positions of Dirac points on the $k_z$ axis of $\GaAsBix$. The magnitude of negative band gap corresponds to band inversion strength of anions.}
\label{tab:dirac_point} 
\end{table}

\section{Mirror-symmetry protected nodal lines in type-B triple-point semimetal $\GaAsBi$ \label{app:nodal}}

\begin{figure}[htb]
\centering
\includegraphics[width=200pt]{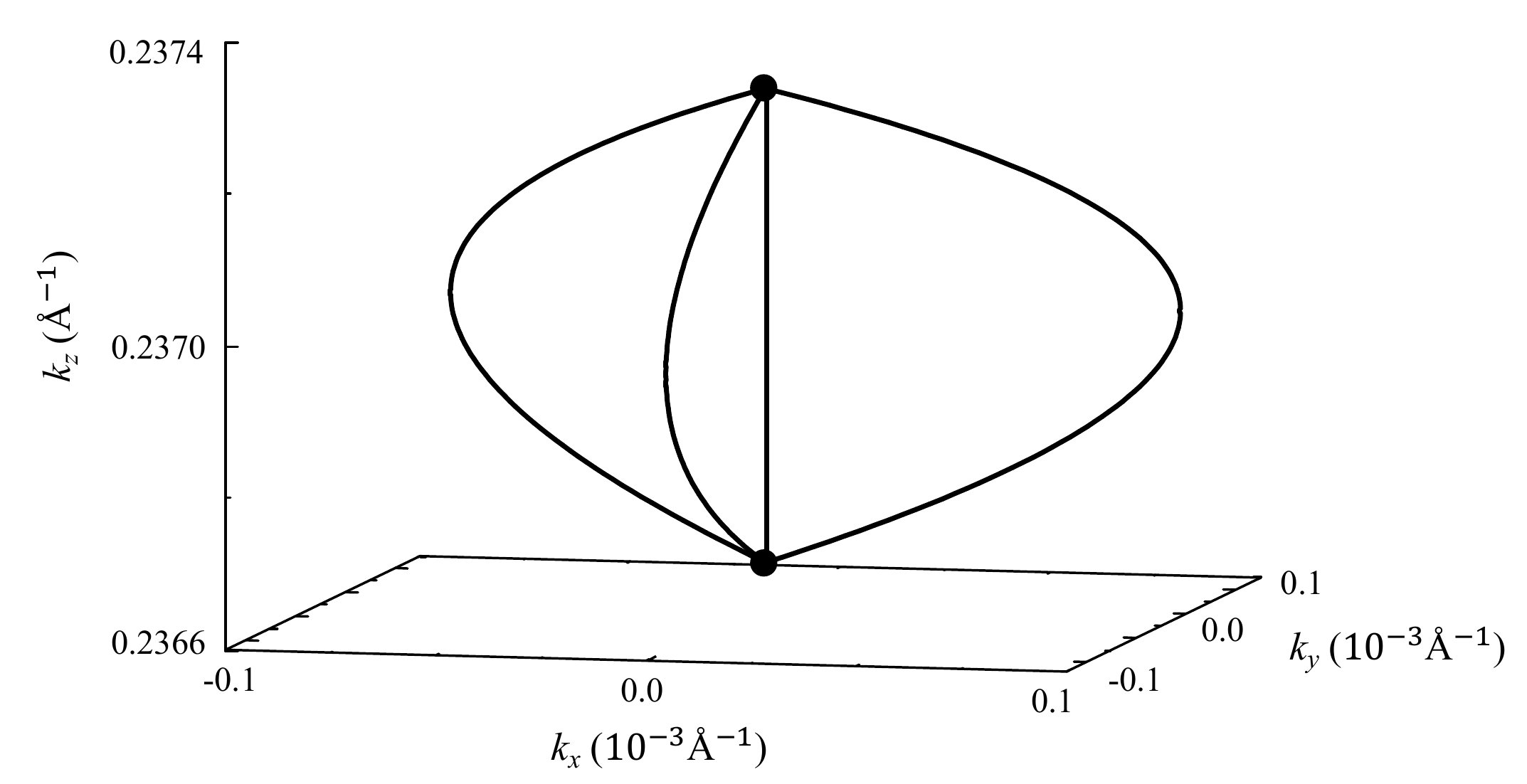}
\caption{The distribution pattern of the mirror-symmetry protected nodal lines in $\GaAsBi$. Each black dot represents one triple-point.}
\label{figa4}
\end{figure}

Since $\GaAsBi$ belongs to the space group No. 160, the triple-points on the Fermi level should be classified into type-B, according to Ref.~\onlinecite{Zhu:2016cb}. Each pair of the type-B triple-points are accompanied by the existence of four nondegenerate nodal lines which are protected by mirror symmetry, as is shown in Figure~\ref{figa4}. Type-B triple-points are different from type-A not only in the number of nodal lines, but also in the Berry phase associated with these nodal lines; the Berry phase accumulated along the loop enclosing these nodal lines are all quantized to be $\phi_B = \pi$.

\bibliographystyle{apsrev4-1}
\bibliography{III_V_wurtzite}

\begin{thebibliography}{44}%
\makeatletter
\providecommand \@ifxundefined [1]{%
 \@ifx{#1\undefined}
}%
\providecommand \@ifnum [1]{%
 \ifnum #1\expandafter \@firstoftwo
 \else \expandafter \@secondoftwo
 \fi
}%
\providecommand \@ifx [1]{%
 \ifx #1\expandafter \@firstoftwo
 \else \expandafter \@secondoftwo
 \fi
}%
\providecommand \natexlab [1]{#1}%
\providecommand \enquote  [1]{``#1''}%
\providecommand \bibnamefont  [1]{#1}%
\providecommand \bibfnamefont [1]{#1}%
\providecommand \citenamefont [1]{#1}%
\providecommand \href@noop [0]{\@secondoftwo}%
\providecommand \href [0]{\begingroup \@sanitize@url \@href}%
\providecommand \@href[1]{\@@startlink{#1}\@@href}%
\providecommand \@@href[1]{\endgroup#1\@@endlink}%
\providecommand \@sanitize@url [0]{\catcode `\\12\catcode `\$12\catcode
  `\&12\catcode `\#12\catcode `\^12\catcode `\_12\catcode `\%12\relax}%
\providecommand \@@startlink[1]{}%
\providecommand \@@endlink[0]{}%
\providecommand \url  [0]{\begingroup\@sanitize@url \@url }%
\providecommand \@url [1]{\endgroup\@href {#1}{\urlprefix }}%
\providecommand \urlprefix  [0]{URL }%
\providecommand \Eprint [0]{\href }%
\providecommand \doibase [0]{http://dx.doi.org/}%
\providecommand \selectlanguage [0]{\@gobble}%
\providecommand \bibinfo  [0]{\@secondoftwo}%
\providecommand \bibfield  [0]{\@secondoftwo}%
\providecommand \translation [1]{[#1]}%
\providecommand \BibitemOpen [0]{}%
\providecommand \bibitemStop [0]{}%
\providecommand \bibitemNoStop [0]{.\EOS\space}%
\providecommand \EOS [0]{\spacefactor3000\relax}%
\providecommand \BibitemShut  [1]{\csname bibitem#1\endcsname}%
\let\auto@bib@innerbib\@empty
\bibitem [{\citenamefont {Armitage}\ \emph {et~al.}(2018)\citenamefont
  {Armitage}, \citenamefont {Mele},\ and\ \citenamefont
  {Vishwanath}}]{Armitage:2018dg}%
  \BibitemOpen
  \bibfield  {author} {\bibinfo {author} {\bibfnamefont {N.~P.}\ \bibnamefont
  {Armitage}}, \bibinfo {author} {\bibfnamefont {E.~J.}\ \bibnamefont {Mele}},
  \ and\ \bibinfo {author} {\bibfnamefont {A.}~\bibnamefont {Vishwanath}},\
  }\href@noop {} {\bibfield  {journal} {\bibinfo  {journal} {Reviews of Modern
  Physics}\ }\textbf {\bibinfo {volume} {90}},\ \bibinfo {pages} {015001}
  (\bibinfo {year} {2018})}\BibitemShut {NoStop}%
\bibitem [{\citenamefont {Young}\ \emph {et~al.}(2012)\citenamefont {Young},
  \citenamefont {Zaheer}, \citenamefont {Teo}, \citenamefont {Kane},
  \citenamefont {Mele},\ and\ \citenamefont {Rappe}}]{Young:2012kz}%
  \BibitemOpen
  \bibfield  {author} {\bibinfo {author} {\bibfnamefont {S.~M.}\ \bibnamefont
  {Young}}, \bibinfo {author} {\bibfnamefont {S.}~\bibnamefont {Zaheer}},
  \bibinfo {author} {\bibfnamefont {J.~C.~Y.}\ \bibnamefont {Teo}}, \bibinfo
  {author} {\bibfnamefont {C.~L.}\ \bibnamefont {Kane}}, \bibinfo {author}
  {\bibfnamefont {E.~J.}\ \bibnamefont {Mele}}, \ and\ \bibinfo {author}
  {\bibfnamefont {A.~M.}\ \bibnamefont {Rappe}},\ }\href@noop {} {\bibfield
  {journal} {\bibinfo  {journal} {Physical Review Letters}\ }\textbf {\bibinfo
  {volume} {108}},\ \bibinfo {pages} {140405} (\bibinfo {year}
  {2012})}\BibitemShut {NoStop}%
\bibitem [{\citenamefont {Wang}\ \emph {et~al.}(2012)\citenamefont {Wang},
  \citenamefont {Sun}, \citenamefont {Chen}, \citenamefont {Franchini},
  \citenamefont {Xu}, \citenamefont {Weng}, \citenamefont {Dai},\ and\
  \citenamefont {Fang}}]{Wang:2012ds}%
  \BibitemOpen
  \bibfield  {author} {\bibinfo {author} {\bibfnamefont {Z.}~\bibnamefont
  {Wang}}, \bibinfo {author} {\bibfnamefont {Y.}~\bibnamefont {Sun}}, \bibinfo
  {author} {\bibfnamefont {X.-Q.}\ \bibnamefont {Chen}}, \bibinfo {author}
  {\bibfnamefont {C.}~\bibnamefont {Franchini}}, \bibinfo {author}
  {\bibfnamefont {G.}~\bibnamefont {Xu}}, \bibinfo {author} {\bibfnamefont
  {H.}~\bibnamefont {Weng}}, \bibinfo {author} {\bibfnamefont {X.}~\bibnamefont
  {Dai}}, \ and\ \bibinfo {author} {\bibfnamefont {Z.}~\bibnamefont {Fang}},\
  }\href@noop {} {\bibfield  {journal} {\bibinfo  {journal} {Physical Review
  B}\ }\textbf {\bibinfo {volume} {85}},\ \bibinfo {pages} {323} (\bibinfo
  {year} {2012})}\BibitemShut {NoStop}%
\bibitem [{\citenamefont {Gao}\ \emph {et~al.}(2019)\citenamefont {Gao},
  \citenamefont {Venderbos}, \citenamefont {Kim},\ and\ \citenamefont
  {Rappe}}]{Heng:2019kd}%
  \BibitemOpen
  \bibfield  {author} {\bibinfo {author} {\bibfnamefont {H.}~\bibnamefont
  {Gao}}, \bibinfo {author} {\bibfnamefont {J.~W.~F.}\ \bibnamefont
  {Venderbos}}, \bibinfo {author} {\bibfnamefont {Y.}~\bibnamefont {Kim}}, \
  and\ \bibinfo {author} {\bibfnamefont {A.~M.}\ \bibnamefont {Rappe}},\
  }\href@noop {} {\bibfield  {journal} {\bibinfo  {journal} {Annu. Rev. Mater.
  Res.}\ }\textbf {\bibinfo {volume} {49}},\ \bibinfo {pages} {153} (\bibinfo
  {year} {2019})}\BibitemShut {NoStop}%
\bibitem [{\citenamefont {Borisenko}\ \emph {et~al.}(2014)\citenamefont
  {Borisenko}, \citenamefont {Gibson}, \citenamefont {Evtushinsky},
  \citenamefont {Zabolotnyy}, \citenamefont {B{\"u}chner},\ and\ \citenamefont
  {Cava}}]{Borisenko:2014ed}%
  \BibitemOpen
  \bibfield  {author} {\bibinfo {author} {\bibfnamefont {S.}~\bibnamefont
  {Borisenko}}, \bibinfo {author} {\bibfnamefont {Q.}~\bibnamefont {Gibson}},
  \bibinfo {author} {\bibfnamefont {D.}~\bibnamefont {Evtushinsky}}, \bibinfo
  {author} {\bibfnamefont {V.}~\bibnamefont {Zabolotnyy}}, \bibinfo {author}
  {\bibfnamefont {B.}~\bibnamefont {B{\"u}chner}}, \ and\ \bibinfo {author}
  {\bibfnamefont {R.~J.}\ \bibnamefont {Cava}},\ }\href@noop {} {\bibfield
  {journal} {\bibinfo  {journal} {Physical Review Letters}\ }\textbf {\bibinfo
  {volume} {113}},\ \bibinfo {pages} {027603} (\bibinfo {year}
  {2014})}\BibitemShut {NoStop}%
\bibitem [{\citenamefont {Burkov}(2016)}]{Burkov:2016hj}%
  \BibitemOpen
  \bibfield  {author} {\bibinfo {author} {\bibfnamefont {A.~A.}\ \bibnamefont
  {Burkov}},\ }\href@noop {} {\bibfield  {journal} {\bibinfo  {journal} {Nature
  Materials}\ }\textbf {\bibinfo {volume} {15}},\ \bibinfo {pages} {1145}
  (\bibinfo {year} {2016})}\BibitemShut {NoStop}%
\bibitem [{\citenamefont {Osterhoudt}\ \emph {et~al.}(2019)\citenamefont
  {Osterhoudt}, \citenamefont {Diebel}, \citenamefont {Gray}, \citenamefont
  {Yang}, \citenamefont {Stanco}, \citenamefont {Huang}, \citenamefont {Shen},
  \citenamefont {Ni}, \citenamefont {Moll}, \citenamefont {Ran},\ and\
  \citenamefont {Burch}}]{Osterhoudt:2019ju}%
  \BibitemOpen
  \bibfield  {author} {\bibinfo {author} {\bibfnamefont {G.~B.}\ \bibnamefont
  {Osterhoudt}}, \bibinfo {author} {\bibfnamefont {L.~K.}\ \bibnamefont
  {Diebel}}, \bibinfo {author} {\bibfnamefont {M.~J.}\ \bibnamefont {Gray}},
  \bibinfo {author} {\bibfnamefont {X.}~\bibnamefont {Yang}}, \bibinfo {author}
  {\bibfnamefont {J.}~\bibnamefont {Stanco}}, \bibinfo {author} {\bibfnamefont
  {X.}~\bibnamefont {Huang}}, \bibinfo {author} {\bibfnamefont
  {B.}~\bibnamefont {Shen}}, \bibinfo {author} {\bibfnamefont {N.}~\bibnamefont
  {Ni}}, \bibinfo {author} {\bibfnamefont {P.~J.~W.}\ \bibnamefont {Moll}},
  \bibinfo {author} {\bibfnamefont {Y.}~\bibnamefont {Ran}}, \ and\ \bibinfo
  {author} {\bibfnamefont {K.~S.}\ \bibnamefont {Burch}},\ }\href@noop {}
  {\bibfield  {journal} {\bibinfo  {journal} {Nature Materials}\ ,\ \bibinfo
  {pages} {1}} (\bibinfo {year} {2019})}\BibitemShut {NoStop}%
\bibitem [{\citenamefont {Zhu}\ \emph {et~al.}(2017)\citenamefont {Zhu},
  \citenamefont {Meng}, \citenamefont {Yuan}, \citenamefont {Luo},
  \citenamefont {Wang}, \citenamefont {Lv}, \citenamefont {He}, \citenamefont
  {Xu}, \citenamefont {Liu}, \citenamefont {Zhang}, \citenamefont {Shi},
  \citenamefont {Zhang}, \citenamefont {Zhu}, \citenamefont {Wang},
  \citenamefont {Xiu},\ and\ \citenamefont {Li}}]{Zhu:2017hf}%
  \BibitemOpen
  \bibfield  {author} {\bibinfo {author} {\bibfnamefont {C.}~\bibnamefont
  {Zhu}}, \bibinfo {author} {\bibfnamefont {Y.}~\bibnamefont {Meng}}, \bibinfo
  {author} {\bibfnamefont {X.}~\bibnamefont {Yuan}}, \bibinfo {author}
  {\bibfnamefont {H.}~\bibnamefont {Luo}}, \bibinfo {author} {\bibfnamefont
  {Y.}~\bibnamefont {Wang}}, \bibinfo {author} {\bibfnamefont {X.}~\bibnamefont
  {Lv}}, \bibinfo {author} {\bibfnamefont {L.}~\bibnamefont {He}}, \bibinfo
  {author} {\bibfnamefont {Y.}~\bibnamefont {Xu}}, \bibinfo {author}
  {\bibfnamefont {J.}~\bibnamefont {Liu}}, \bibinfo {author} {\bibfnamefont
  {C.}~\bibnamefont {Zhang}}, \bibinfo {author} {\bibfnamefont
  {Y.}~\bibnamefont {Shi}}, \bibinfo {author} {\bibfnamefont {R.}~\bibnamefont
  {Zhang}}, \bibinfo {author} {\bibfnamefont {S.}~\bibnamefont {Zhu}}, \bibinfo
  {author} {\bibfnamefont {F.}~\bibnamefont {Wang}}, \bibinfo {author}
  {\bibfnamefont {F.}~\bibnamefont {Xiu}}, \ and\ \bibinfo {author}
  {\bibfnamefont {J.}~\bibnamefont {Li}},\ }\href@noop {} {\bibfield  {journal}
  {\bibinfo  {journal} {Nature Communications}\ }\textbf {\bibinfo {volume}
  {8}},\ \bibinfo {pages} {1} (\bibinfo {year} {2017})}\BibitemShut {NoStop}%
\bibitem [{\citenamefont {Liu}\ \emph {et~al.}(2018)\citenamefont {Liu},
  \citenamefont {Liu}, \citenamefont {Wang}, \citenamefont {He}, \citenamefont
  {Wan}, \citenamefont {Xu}, \citenamefont {Shi}, \citenamefont {Zhang},\ and\
  \citenamefont {Wang}}]{Liu:2018du}%
  \BibitemOpen
  \bibfield  {author} {\bibinfo {author} {\bibfnamefont {Y.}~\bibnamefont
  {Liu}}, \bibinfo {author} {\bibfnamefont {C.}~\bibnamefont {Liu}}, \bibinfo
  {author} {\bibfnamefont {X.}~\bibnamefont {Wang}}, \bibinfo {author}
  {\bibfnamefont {L.}~\bibnamefont {He}}, \bibinfo {author} {\bibfnamefont
  {X.}~\bibnamefont {Wan}}, \bibinfo {author} {\bibfnamefont {Y.}~\bibnamefont
  {Xu}}, \bibinfo {author} {\bibfnamefont {Y.}~\bibnamefont {Shi}}, \bibinfo
  {author} {\bibfnamefont {R.}~\bibnamefont {Zhang}}, \ and\ \bibinfo {author}
  {\bibfnamefont {F.}~\bibnamefont {Wang}},\ }\href@noop {} {\bibfield
  {journal} {\bibinfo  {journal} {Scientific Reports}\ ,\ \bibinfo {pages} {1}}
  (\bibinfo {year} {2018})}\BibitemShut {NoStop}%
\bibitem [{\citenamefont {Norton}(2002)}]{Norton:2002}%
  \BibitemOpen
  \bibfield  {author} {\bibinfo {author} {\bibfnamefont {P.}~\bibnamefont
  {Norton}},\ }\href@noop {} {\bibfield  {journal} {\bibinfo  {journal}
  {Optoelectronics Review}\ }\textbf {\bibinfo {volume} {10}},\ \bibinfo
  {pages} {159} (\bibinfo {year} {2002})}\BibitemShut {NoStop}%
\bibitem [{\citenamefont {Chan}\ \emph {et~al.}(2017)\citenamefont {Chan},
  \citenamefont {Lindner}, \citenamefont {Refael},\ and\ \citenamefont
  {Lee}}]{Chan:2017ef}%
  \BibitemOpen
  \bibfield  {author} {\bibinfo {author} {\bibfnamefont {C.-K.}\ \bibnamefont
  {Chan}}, \bibinfo {author} {\bibfnamefont {N.~H.}\ \bibnamefont {Lindner}},
  \bibinfo {author} {\bibfnamefont {G.}~\bibnamefont {Refael}}, \ and\ \bibinfo
  {author} {\bibfnamefont {P.~A.}\ \bibnamefont {Lee}},\ }\href@noop {}
  {\bibfield  {journal} {\bibinfo  {journal} {Physical Review B}\ }\textbf
  {\bibinfo {volume} {95}},\ \bibinfo {pages} {041104} (\bibinfo {year}
  {2017})}\BibitemShut {NoStop}%
\bibitem [{\citenamefont {Vazifeh}\ and\ \citenamefont
  {Franz}(2013)}]{Vazifeh:2013fe}%
  \BibitemOpen
  \bibfield  {author} {\bibinfo {author} {\bibfnamefont {M.~M.}\ \bibnamefont
  {Vazifeh}}\ and\ \bibinfo {author} {\bibfnamefont {M.}~\bibnamefont
  {Franz}},\ }\href@noop {} {\bibfield  {journal} {\bibinfo  {journal}
  {Physical Review Letters}\ }\textbf {\bibinfo {volume} {111}},\ \bibinfo
  {pages} {027201} (\bibinfo {year} {2013})}\BibitemShut {NoStop}%
\bibitem [{\citenamefont {Zhong}\ \emph {et~al.}(2016)\citenamefont {Zhong},
  \citenamefont {Moore},\ and\ \citenamefont {Souza}}]{Zhong:2016dja}%
  \BibitemOpen
  \bibfield  {author} {\bibinfo {author} {\bibfnamefont {S.}~\bibnamefont
  {Zhong}}, \bibinfo {author} {\bibfnamefont {J.~E.}\ \bibnamefont {Moore}}, \
  and\ \bibinfo {author} {\bibfnamefont {I.}~\bibnamefont {Souza}},\
  }\href@noop {} {\bibfield  {journal} {\bibinfo  {journal} {Physical Review
  Letters}\ }\textbf {\bibinfo {volume} {116}},\ \bibinfo {pages} {677}
  (\bibinfo {year} {2016})}\BibitemShut {NoStop}%
\bibitem [{\citenamefont {Mclver}\ \emph {et~al.}(2012)\citenamefont {Mclver},
  \citenamefont {Hsieh}, \citenamefont {Drapcho}, \citenamefont {Torchinsky},
  \citenamefont {Gardner}, \citenamefont {Lee},\ and\ \citenamefont
  {Gedik}}]{McIver:2012jp}%
  \BibitemOpen
  \bibfield  {author} {\bibinfo {author} {\bibfnamefont {J.~W.}\ \bibnamefont
  {Mclver}}, \bibinfo {author} {\bibfnamefont {D.}~\bibnamefont {Hsieh}},
  \bibinfo {author} {\bibfnamefont {S.~G.}\ \bibnamefont {Drapcho}}, \bibinfo
  {author} {\bibfnamefont {D.~H.}\ \bibnamefont {Torchinsky}}, \bibinfo
  {author} {\bibfnamefont {D.~R.}\ \bibnamefont {Gardner}}, \bibinfo {author}
  {\bibfnamefont {Y.~S.}\ \bibnamefont {Lee}}, \ and\ \bibinfo {author}
  {\bibfnamefont {N.}~\bibnamefont {Gedik}},\ }\href@noop {} {\bibfield
  {journal} {\bibinfo  {journal} {Phys. Rev. B}\ }\textbf {\bibinfo {volume}
  {86}} (\bibinfo {year} {2012})}\BibitemShut {NoStop}%
\bibitem [{\citenamefont {Shi}\ \emph {et~al.}(2016)\citenamefont {Shi},
  \citenamefont {Zhanng}, \citenamefont {Yao}, \citenamefont {Ji},
  \citenamefont {Qian}, \citenamefont {Qiao}, \citenamefont {Shen},\ and\
  \citenamefont {Liu}}]{Shi:2016yw}%
  \BibitemOpen
  \bibfield  {author} {\bibinfo {author} {\bibfnamefont {H.}~\bibnamefont
  {Shi}}, \bibinfo {author} {\bibfnamefont {Y.}~\bibnamefont {Zhanng}},
  \bibinfo {author} {\bibfnamefont {M.-Y.}\ \bibnamefont {Yao}}, \bibinfo
  {author} {\bibfnamefont {F.-H.}\ \bibnamefont {Ji}}, \bibinfo {author}
  {\bibfnamefont {D.}~\bibnamefont {Qian}}, \bibinfo {author} {\bibfnamefont
  {S.}~\bibnamefont {Qiao}}, \bibinfo {author} {\bibfnamefont {Y.-R.}\
  \bibnamefont {Shen}}, \ and\ \bibinfo {author} {\bibfnamefont {W.-T.}\
  \bibnamefont {Liu}},\ }\href@noop {} {\bibfield  {journal} {\bibinfo
  {journal} {Phys. Rev. B}\ }\textbf {\bibinfo {volume} {94}} (\bibinfo {year}
  {2016})}\BibitemShut {NoStop}%
\bibitem [{\citenamefont {Kharzeev}(2014)}]{Kharzeev:2014fra}%
  \BibitemOpen
  \bibfield  {author} {\bibinfo {author} {\bibfnamefont {D.~E.}\ \bibnamefont
  {Kharzeev}},\ }\href@noop {} {\bibfield  {journal} {\bibinfo  {journal}
  {Progress in Particle and Nuclear Physics}\ }\textbf {\bibinfo {volume}
  {75}},\ \bibinfo {pages} {133} (\bibinfo {year} {2014})}\BibitemShut
  {NoStop}%
\bibitem [{\citenamefont {Son}\ and\ \citenamefont
  {Spivak}(2013)}]{Son:2013jza}%
  \BibitemOpen
  \bibfield  {author} {\bibinfo {author} {\bibfnamefont {D.~T.}\ \bibnamefont
  {Son}}\ and\ \bibinfo {author} {\bibfnamefont {B.~Z.}\ \bibnamefont
  {Spivak}},\ }\href@noop {} {\bibfield  {journal} {\bibinfo  {journal}
  {Physical Review B}\ }\textbf {\bibinfo {volume} {88}},\ \bibinfo {pages}
  {390} (\bibinfo {year} {2013})}\BibitemShut {NoStop}%
\bibitem [{\citenamefont {Son}\ and\ \citenamefont
  {Yamamoto}(2012)}]{Son:2012eoa}%
  \BibitemOpen
  \bibfield  {author} {\bibinfo {author} {\bibfnamefont {D.~T.}\ \bibnamefont
  {Son}}\ and\ \bibinfo {author} {\bibfnamefont {N.}~\bibnamefont {Yamamoto}},\
  }\href@noop {} {\bibfield  {journal} {\bibinfo  {journal} {Physical Review
  Letters}\ }\textbf {\bibinfo {volume} {109}},\ \bibinfo {pages} {920}
  (\bibinfo {year} {2012})}\BibitemShut {NoStop}%
\bibitem [{\citenamefont {Huang}\ \emph {et~al.}(2014)\citenamefont {Huang},
  \citenamefont {Liu},\ and\ \citenamefont {Duan}}]{Huang:2014hg}%
  \BibitemOpen
  \bibfield  {author} {\bibinfo {author} {\bibfnamefont {H.}~\bibnamefont
  {Huang}}, \bibinfo {author} {\bibfnamefont {J.}~\bibnamefont {Liu}}, \ and\
  \bibinfo {author} {\bibfnamefont {W.}~\bibnamefont {Duan}},\ }\href@noop {}
  {\bibfield  {journal} {\bibinfo  {journal} {Physical Review B}\ }\textbf
  {\bibinfo {volume} {90}},\ \bibinfo {pages} {195105} (\bibinfo {year}
  {2014})}\BibitemShut {NoStop}%
\bibitem [{\citenamefont {Bernevig}\ \emph {et~al.}(2006)\citenamefont
  {Bernevig}, \citenamefont {Hughes},\ and\ \citenamefont
  {Zhang}}]{Bernvig:2006xw}%
  \BibitemOpen
  \bibfield  {author} {\bibinfo {author} {\bibfnamefont {B.~A.}\ \bibnamefont
  {Bernevig}}, \bibinfo {author} {\bibfnamefont {T.~L.}\ \bibnamefont
  {Hughes}}, \ and\ \bibinfo {author} {\bibfnamefont {S.-C.}\ \bibnamefont
  {Zhang}},\ }\href@noop {} {\bibfield  {journal} {\bibinfo  {journal}
  {Science}\ }\textbf {\bibinfo {volume} {314}},\ \bibinfo {pages} {1757}
  (\bibinfo {year} {2006})}\BibitemShut {NoStop}%
\bibitem [{\citenamefont {Winkler}\ \emph {et~al.}(2016)\citenamefont
  {Winkler}, \citenamefont {Wu}, \citenamefont {Troyer}, \citenamefont
  {Krogstrup},\ and\ \citenamefont {Soluyanov}}]{Winkler:2016kk}%
  \BibitemOpen
  \bibfield  {author} {\bibinfo {author} {\bibfnamefont {G.~W.}\ \bibnamefont
  {Winkler}}, \bibinfo {author} {\bibfnamefont {Q.}~\bibnamefont {Wu}},
  \bibinfo {author} {\bibfnamefont {M.}~\bibnamefont {Troyer}}, \bibinfo
  {author} {\bibfnamefont {P.}~\bibnamefont {Krogstrup}}, \ and\ \bibinfo
  {author} {\bibfnamefont {A.~A.}\ \bibnamefont {Soluyanov}},\ }\href@noop {}
  {\bibfield  {journal} {\bibinfo  {journal} {Physical Review Letters}\
  }\textbf {\bibinfo {volume} {117}},\ \bibinfo {pages} {78} (\bibinfo {year}
  {2016})}\BibitemShut {NoStop}%
\bibitem [{\citenamefont {Collino}\ \emph {et~al.}(2011)\citenamefont
  {Collino}, \citenamefont {Wood}, \citenamefont {Estrada}, \citenamefont
  {Dick}, \citenamefont {Ro}, \citenamefont {Soles}, \citenamefont {Wang},
  \citenamefont {Thouless},\ and\ \citenamefont {Goldman}}]{Collino:2011er}%
  \BibitemOpen
  \bibfield  {author} {\bibinfo {author} {\bibfnamefont {R.~R.}\ \bibnamefont
  {Collino}}, \bibinfo {author} {\bibfnamefont {A.~W.}\ \bibnamefont {Wood}},
  \bibinfo {author} {\bibfnamefont {N.~M.}\ \bibnamefont {Estrada}}, \bibinfo
  {author} {\bibfnamefont {B.~B.}\ \bibnamefont {Dick}}, \bibinfo {author}
  {\bibfnamefont {H.~W.}\ \bibnamefont {Ro}}, \bibinfo {author} {\bibfnamefont
  {C.~L.}\ \bibnamefont {Soles}}, \bibinfo {author} {\bibfnamefont {Y.~Q.}\
  \bibnamefont {Wang}}, \bibinfo {author} {\bibfnamefont {M.~D.}\ \bibnamefont
  {Thouless}}, \ and\ \bibinfo {author} {\bibfnamefont {R.~S.}\ \bibnamefont
  {Goldman}},\ }\href@noop {} {\bibfield  {journal} {\bibinfo  {journal}
  {Journal of Vacuum Science {\&} Technology A: Vacuum, Surfaces, and Films}\
  }\textbf {\bibinfo {volume} {29}},\ \bibinfo {pages} {060601} (\bibinfo
  {year} {2011})}\BibitemShut {NoStop}%
\bibitem [{\citenamefont {Wood}\ \emph {et~al.}(2011)\citenamefont {Wood},
  \citenamefont {Weng}, \citenamefont {Wang},\ and\ \citenamefont
  {Goldman}}]{Wood:2011ec}%
  \BibitemOpen
  \bibfield  {author} {\bibinfo {author} {\bibfnamefont {A.~W.}\ \bibnamefont
  {Wood}}, \bibinfo {author} {\bibfnamefont {X.}~\bibnamefont {Weng}}, \bibinfo
  {author} {\bibfnamefont {Y.~Q.}\ \bibnamefont {Wang}}, \ and\ \bibinfo
  {author} {\bibfnamefont {R.~S.}\ \bibnamefont {Goldman}},\ }\href@noop {}
  {\bibfield  {journal} {\bibinfo  {journal} {Applied Physics Letters}\
  }\textbf {\bibinfo {volume} {99}},\ \bibinfo {pages} {093108} (\bibinfo
  {year} {2011})}\BibitemShut {NoStop}%
\bibitem [{\citenamefont {Wood}\ \emph {et~al.}(2012)\citenamefont {Wood},
  \citenamefont {Collino}, \citenamefont {Wang}, \citenamefont {Wang},\ and\
  \citenamefont {Goldman}}]{Wood:2012bs}%
  \BibitemOpen
  \bibfield  {author} {\bibinfo {author} {\bibfnamefont {A.~W.}\ \bibnamefont
  {Wood}}, \bibinfo {author} {\bibfnamefont {R.~R.}\ \bibnamefont {Collino}},
  \bibinfo {author} {\bibfnamefont {P.~T.}\ \bibnamefont {Wang}}, \bibinfo
  {author} {\bibfnamefont {Y.~Q.}\ \bibnamefont {Wang}}, \ and\ \bibinfo
  {author} {\bibfnamefont {R.~S.}\ \bibnamefont {Goldman}},\ }\href@noop {}
  {\bibfield  {journal} {\bibinfo  {journal} {Applied Physics Letters}\
  }\textbf {\bibinfo {volume} {100}},\ \bibinfo {pages} {203113} (\bibinfo
  {year} {2012})}\BibitemShut {NoStop}%
\bibitem [{\citenamefont {Gao}\ \emph {et~al.}(2018)\citenamefont {Gao},
  \citenamefont {Kim}, \citenamefont {Venderbos}, \citenamefont {Kane},
  \citenamefont {Mele}, \citenamefont {Rappe},\ and\ \citenamefont
  {Ren}}]{Gao:2018bu}%
  \BibitemOpen
  \bibfield  {author} {\bibinfo {author} {\bibfnamefont {H.}~\bibnamefont
  {Gao}}, \bibinfo {author} {\bibfnamefont {Y.}~\bibnamefont {Kim}}, \bibinfo
  {author} {\bibfnamefont {J.~W.~F.}\ \bibnamefont {Venderbos}}, \bibinfo
  {author} {\bibfnamefont {C.~L.}\ \bibnamefont {Kane}}, \bibinfo {author}
  {\bibfnamefont {E.~J.}\ \bibnamefont {Mele}}, \bibinfo {author}
  {\bibfnamefont {A.~M.}\ \bibnamefont {Rappe}}, \ and\ \bibinfo {author}
  {\bibfnamefont {W.}~\bibnamefont {Ren}},\ }\href@noop {} {\bibfield
  {journal} {\bibinfo  {journal} {Physical Review Letters}\ }\textbf {\bibinfo
  {volume} {121}},\ \bibinfo {pages} {106404} (\bibinfo {year}
  {2018})}\BibitemShut {NoStop}%
\bibitem [{\citenamefont {Chen}\ \emph {et~al.}(2017)\citenamefont {Chen},
  \citenamefont {Wang}, \citenamefont {Liu}, \citenamefont {Yu}, \citenamefont
  {Sheng}, \citenamefont {Chen},\ and\ \citenamefont {Yang}}]{Chen:2017gg}%
  \BibitemOpen
  \bibfield  {author} {\bibinfo {author} {\bibfnamefont {C.}~\bibnamefont
  {Chen}}, \bibinfo {author} {\bibfnamefont {S.-S.}\ \bibnamefont {Wang}},
  \bibinfo {author} {\bibfnamefont {L.}~\bibnamefont {Liu}}, \bibinfo {author}
  {\bibfnamefont {Z.-M.}\ \bibnamefont {Yu}}, \bibinfo {author} {\bibfnamefont
  {X.-L.}\ \bibnamefont {Sheng}}, \bibinfo {author} {\bibfnamefont
  {Z.}~\bibnamefont {Chen}}, \ and\ \bibinfo {author} {\bibfnamefont {S.~A.}\
  \bibnamefont {Yang}},\ }\href@noop {} {\bibfield  {journal} {\bibinfo
  {journal} {Physical Review Materials}\ }\textbf {\bibinfo {volume} {1}},\
  \bibinfo {pages} {044201} (\bibinfo {year} {2017})}\BibitemShut {NoStop}%
\bibitem [{\citenamefont {Vurgaftman}\ \emph {et~al.}(2001)\citenamefont
  {Vurgaftman}, \citenamefont {Meyer},\ and\ \citenamefont
  {Ram-Mohan}}]{Vurgaftman:2001bu}%
  \BibitemOpen
  \bibfield  {author} {\bibinfo {author} {\bibfnamefont {I.}~\bibnamefont
  {Vurgaftman}}, \bibinfo {author} {\bibfnamefont {J.~R.}\ \bibnamefont
  {Meyer}}, \ and\ \bibinfo {author} {\bibfnamefont {L.~R.}\ \bibnamefont
  {Ram-Mohan}},\ }\href@noop {} {\bibfield  {journal} {\bibinfo  {journal}
  {Journal of Applied Physics}\ }\textbf {\bibinfo {volume} {89}},\ \bibinfo
  {pages} {5815} (\bibinfo {year} {2001})}\BibitemShut {NoStop}%
\bibitem [{\citenamefont {Zhang}\ \emph {et~al.}(2013)\citenamefont {Zhang},
  \citenamefont {Ma}, \citenamefont {Liu}, \citenamefont {Xu}, \citenamefont
  {Liu}, \citenamefont {Liu}, \citenamefont {Liu}, \citenamefont {Wang},\ and\
  \citenamefont {Wu}}]{Zhang:2013ea}%
  \BibitemOpen
  \bibfield  {author} {\bibinfo {author} {\bibfnamefont {X.~M.}\ \bibnamefont
  {Zhang}}, \bibinfo {author} {\bibfnamefont {R.~S.}\ \bibnamefont {Ma}},
  \bibinfo {author} {\bibfnamefont {X.~C.}\ \bibnamefont {Liu}}, \bibinfo
  {author} {\bibfnamefont {G.~Z.}\ \bibnamefont {Xu}}, \bibinfo {author}
  {\bibfnamefont {E.~K.}\ \bibnamefont {Liu}}, \bibinfo {author} {\bibfnamefont
  {G.~D.}\ \bibnamefont {Liu}}, \bibinfo {author} {\bibfnamefont {Z.~Y.}\
  \bibnamefont {Liu}}, \bibinfo {author} {\bibfnamefont {W.~H.}\ \bibnamefont
  {Wang}}, \ and\ \bibinfo {author} {\bibfnamefont {G.~H.}\ \bibnamefont
  {Wu}},\ }\href@noop {} {\bibfield  {journal} {\bibinfo  {journal} {EPL
  (Europhysics Letters)}\ }\textbf {\bibinfo {volume} {103}},\ \bibinfo {pages}
  {57012} (\bibinfo {year} {2013})}\BibitemShut {NoStop}%
\bibitem [{\citenamefont {Ferhat}\ and\ \citenamefont
  {Zaoui}(2006)}]{Ferhat:2006gq}%
  \BibitemOpen
  \bibfield  {author} {\bibinfo {author} {\bibfnamefont {M.}~\bibnamefont
  {Ferhat}}\ and\ \bibinfo {author} {\bibfnamefont {A.}~\bibnamefont {Zaoui}},\
  }\href@noop {} {\bibfield  {journal} {\bibinfo  {journal} {Physical Review
  B}\ }\textbf {\bibinfo {volume} {73}},\ \bibinfo {pages} {853} (\bibinfo
  {year} {2006})}\BibitemShut {NoStop}%
\bibitem [{\citenamefont {Wang}\ \emph {et~al.}(2017)\citenamefont {Wang},
  \citenamefont {Zhang}, \citenamefont {Yue}, \citenamefont {Liang},
  \citenamefont {Chen}, \citenamefont {Li}, \citenamefont {Lu}, \citenamefont
  {Shao},\ and\ \citenamefont {Wang}}]{Wang:2017kx}%
  \BibitemOpen
  \bibfield  {author} {\bibinfo {author} {\bibfnamefont {L.}~\bibnamefont
  {Wang}}, \bibinfo {author} {\bibfnamefont {L.}~\bibnamefont {Zhang}},
  \bibinfo {author} {\bibfnamefont {L.}~\bibnamefont {Yue}}, \bibinfo {author}
  {\bibfnamefont {D.}~\bibnamefont {Liang}}, \bibinfo {author} {\bibfnamefont
  {X.}~\bibnamefont {Chen}}, \bibinfo {author} {\bibfnamefont {Y.}~\bibnamefont
  {Li}}, \bibinfo {author} {\bibfnamefont {P.}~\bibnamefont {Lu}}, \bibinfo
  {author} {\bibfnamefont {J.}~\bibnamefont {Shao}}, \ and\ \bibinfo {author}
  {\bibfnamefont {S.}~\bibnamefont {Wang}},\ }\href@noop {} {\bibfield
  {journal} {\bibinfo  {journal} {Crystals}\ }\textbf {\bibinfo {volume} {7}},\
  \bibinfo {pages} {63} (\bibinfo {year} {2017})}\BibitemShut {NoStop}%
\bibitem [{\citenamefont {Soluyanov}\ \emph {et~al.}(2015)\citenamefont
  {Soluyanov}, \citenamefont {Gresch}, \citenamefont {Wang}, \citenamefont
  {Wu}, \citenamefont {Troyer}, \citenamefont {Dai},\ and\ \citenamefont
  {Bernevig}}]{Soluyanov:2015cn}%
  \BibitemOpen
  \bibfield  {author} {\bibinfo {author} {\bibfnamefont {A.~A.}\ \bibnamefont
  {Soluyanov}}, \bibinfo {author} {\bibfnamefont {D.}~\bibnamefont {Gresch}},
  \bibinfo {author} {\bibfnamefont {Z.}~\bibnamefont {Wang}}, \bibinfo {author}
  {\bibfnamefont {Q.}~\bibnamefont {Wu}}, \bibinfo {author} {\bibfnamefont
  {M.}~\bibnamefont {Troyer}}, \bibinfo {author} {\bibfnamefont
  {X.}~\bibnamefont {Dai}}, \ and\ \bibinfo {author} {\bibfnamefont {B.~A.}\
  \bibnamefont {Bernevig}},\ }\href@noop {} {\bibfield  {journal} {\bibinfo
  {journal} {Nature}\ }\textbf {\bibinfo {volume} {527}},\ \bibinfo {pages}
  {495} (\bibinfo {year} {2015})}\BibitemShut {NoStop}%
\bibitem [{\citenamefont {Lv}\ \emph {et~al.}(2015)\citenamefont {Lv},
  \citenamefont {Weng}, \citenamefont {Fu}, \citenamefont {Wang}, \citenamefont
  {Miao}, \citenamefont {Ma}, \citenamefont {Richard}, \citenamefont {Huang},
  \citenamefont {Zhao}, \citenamefont {Chen}, \citenamefont {Fang},
  \citenamefont {Dai}, \citenamefont {Qian},\ and\ \citenamefont
  {Ding}}]{Lv:2015yn}%
  \BibitemOpen
  \bibfield  {author} {\bibinfo {author} {\bibfnamefont {B.-Q.}\ \bibnamefont
  {Lv}}, \bibinfo {author} {\bibfnamefont {H.-M.}\ \bibnamefont {Weng}},
  \bibinfo {author} {\bibfnamefont {B.-B.}\ \bibnamefont {Fu}}, \bibinfo
  {author} {\bibfnamefont {X.-P.}\ \bibnamefont {Wang}}, \bibinfo {author}
  {\bibfnamefont {H.}~\bibnamefont {Miao}}, \bibinfo {author} {\bibfnamefont
  {J.}~\bibnamefont {Ma}}, \bibinfo {author} {\bibfnamefont {P.}~\bibnamefont
  {Richard}}, \bibinfo {author} {\bibfnamefont {X.-C.}\ \bibnamefont {Huang}},
  \bibinfo {author} {\bibfnamefont {L.-X.}\ \bibnamefont {Zhao}}, \bibinfo
  {author} {\bibfnamefont {G.-F.}\ \bibnamefont {Chen}}, \bibinfo {author}
  {\bibfnamefont {Z.}~\bibnamefont {Fang}}, \bibinfo {author} {\bibfnamefont
  {X.}~\bibnamefont {Dai}}, \bibinfo {author} {\bibfnamefont {T.}~\bibnamefont
  {Qian}}, \ and\ \bibinfo {author} {\bibfnamefont {H.}~\bibnamefont {Ding}},\
  }\href@noop {} {\bibfield  {journal} {\bibinfo  {journal} {Phys. Rev. X}\
  }\textbf {\bibinfo {volume} {5}} (\bibinfo {year} {2015})}\BibitemShut
  {NoStop}%
\bibitem [{\citenamefont {Jiang}\ \emph {et~al.}(2017)\citenamefont {Jiang},
  \citenamefont {Liu}, \citenamefont {Sun}, \citenamefont {Yang}, \citenamefont
  {Rajamathi}, \citenamefont {Qi}, \citenamefont {Yang}, \citenamefont {Chen},
  \citenamefont {Peng}, \citenamefont {Hwang}, \citenamefont {Sun},
  \citenamefont {Mo}, \citenamefont {Vobornik}, \citenamefont {Fujii},
  \citenamefont {Parkin S. S. P.~annd Felser}, \citenamefont {Yan},\ and\
  \citenamefont {Chen}}]{Jiang:2017ba}%
  \BibitemOpen
  \bibfield  {author} {\bibinfo {author} {\bibfnamefont {J.}~\bibnamefont
  {Jiang}}, \bibinfo {author} {\bibfnamefont {Z.-K.}\ \bibnamefont {Liu}},
  \bibinfo {author} {\bibfnamefont {Y.}~\bibnamefont {Sun}}, \bibinfo {author}
  {\bibfnamefont {H.-F.}\ \bibnamefont {Yang}}, \bibinfo {author}
  {\bibfnamefont {C.~R.}\ \bibnamefont {Rajamathi}}, \bibinfo {author}
  {\bibfnamefont {Y.-P.}\ \bibnamefont {Qi}}, \bibinfo {author} {\bibfnamefont
  {L.-X.}\ \bibnamefont {Yang}}, \bibinfo {author} {\bibfnamefont
  {C.}~\bibnamefont {Chen}}, \bibinfo {author} {\bibfnamefont {H.}~\bibnamefont
  {Peng}}, \bibinfo {author} {\bibfnamefont {C.-C.}\ \bibnamefont {Hwang}},
  \bibinfo {author} {\bibfnamefont {S.-Z.}\ \bibnamefont {Sun}}, \bibinfo
  {author} {\bibfnamefont {S.-K.}\ \bibnamefont {Mo}}, \bibinfo {author}
  {\bibfnamefont {I.}~\bibnamefont {Vobornik}}, \bibinfo {author}
  {\bibfnamefont {J.}~\bibnamefont {Fujii}}, \bibinfo {author} {\bibfnamefont
  {C.}~\bibnamefont {Parkin S. S. P.~annd Felser}}, \bibinfo {author}
  {\bibfnamefont {B.-H.}\ \bibnamefont {Yan}}, \ and\ \bibinfo {author}
  {\bibfnamefont {Y.-L.}\ \bibnamefont {Chen}},\ }\href@noop {} {\bibfield
  {journal} {\bibinfo  {journal} {Nat. Comm.}\ }\textbf {\bibinfo {volume} {8}}
  (\bibinfo {year} {2017})}\BibitemShut {NoStop}%
\bibitem [{\citenamefont {Zhu}\ \emph {et~al.}(2016)\citenamefont {Zhu},
  \citenamefont {Winkler}, \citenamefont {Wu}, \citenamefont {Li},\ and\
  \citenamefont {Soluyanov}}]{Zhu:2016cb}%
  \BibitemOpen
  \bibfield  {author} {\bibinfo {author} {\bibfnamefont {Z.}~\bibnamefont
  {Zhu}}, \bibinfo {author} {\bibfnamefont {G.~W.}\ \bibnamefont {Winkler}},
  \bibinfo {author} {\bibfnamefont {Q.}~\bibnamefont {Wu}}, \bibinfo {author}
  {\bibfnamefont {J.}~\bibnamefont {Li}}, \ and\ \bibinfo {author}
  {\bibfnamefont {A.~A.}\ \bibnamefont {Soluyanov}},\ }\href@noop {} {\bibfield
   {journal} {\bibinfo  {journal} {Physical Review X}\ }\textbf {\bibinfo
  {volume} {6}},\ \bibinfo {pages} {1088} (\bibinfo {year} {2016})}\BibitemShut
  {NoStop}%
\bibitem [{\citenamefont {Giannozzi}\ \emph {et~al.}(2009)\citenamefont
  {Giannozzi}, \citenamefont {Baroni}, \citenamefont {Bonini}, \citenamefont
  {Calandra}, \citenamefont {Car}, \citenamefont {Cavazzoni}, \citenamefont
  {Ceresoli}, \citenamefont {Chiarotti}, \citenamefont {Cococcioni},
  \citenamefont {Dabo} \emph {et~al.}}]{Giannozzi:2009hx}%
  \BibitemOpen
  \bibfield  {author} {\bibinfo {author} {\bibfnamefont {P.}~\bibnamefont
  {Giannozzi}}, \bibinfo {author} {\bibfnamefont {S.}~\bibnamefont {Baroni}},
  \bibinfo {author} {\bibfnamefont {N.}~\bibnamefont {Bonini}}, \bibinfo
  {author} {\bibfnamefont {M.}~\bibnamefont {Calandra}}, \bibinfo {author}
  {\bibfnamefont {R.}~\bibnamefont {Car}}, \bibinfo {author} {\bibfnamefont
  {C.}~\bibnamefont {Cavazzoni}}, \bibinfo {author} {\bibfnamefont
  {D.}~\bibnamefont {Ceresoli}}, \bibinfo {author} {\bibfnamefont {G.~L.}\
  \bibnamefont {Chiarotti}}, \bibinfo {author} {\bibfnamefont {M.}~\bibnamefont
  {Cococcioni}}, \bibinfo {author} {\bibfnamefont {I.}~\bibnamefont {Dabo}},
  \emph {et~al.},\ }\href@noop {} {\bibfield  {journal} {\bibinfo  {journal}
  {Journal of Physics: Condensed Matter}\ }\textbf {\bibinfo {volume} {21}},\
  \bibinfo {pages} {395502} (\bibinfo {year} {2009})}\BibitemShut {NoStop}%
\bibitem [{\citenamefont {Tixier}\ \emph {et~al.}(2003)\citenamefont {Tixier},
  \citenamefont {Adamcyk}, \citenamefont {Tiedje}, \citenamefont {Francoeur},
  \citenamefont {Mascarenhas}, \citenamefont {Wei},\ and\ \citenamefont
  {Schiettekatte}}]{Tixier:2003jc}%
  \BibitemOpen
  \bibfield  {author} {\bibinfo {author} {\bibfnamefont {S.}~\bibnamefont
  {Tixier}}, \bibinfo {author} {\bibfnamefont {M.}~\bibnamefont {Adamcyk}},
  \bibinfo {author} {\bibfnamefont {T.}~\bibnamefont {Tiedje}}, \bibinfo
  {author} {\bibfnamefont {S.}~\bibnamefont {Francoeur}}, \bibinfo {author}
  {\bibfnamefont {A.}~\bibnamefont {Mascarenhas}}, \bibinfo {author}
  {\bibfnamefont {P.}~\bibnamefont {Wei}}, \ and\ \bibinfo {author}
  {\bibfnamefont {F.}~\bibnamefont {Schiettekatte}},\ }\href@noop {} {\bibfield
   {journal} {\bibinfo  {journal} {Applied Physics Letters}\ }\textbf {\bibinfo
  {volume} {82}},\ \bibinfo {pages} {2245} (\bibinfo {year}
  {2003})}\BibitemShut {NoStop}%
\bibitem [{\citenamefont {Shalindar}\ \emph {et~al.}(2016)\citenamefont
  {Shalindar}, \citenamefont {Webster}, \citenamefont {Wilkens}, \citenamefont
  {Alford},\ and\ \citenamefont {Johnson}}]{Shalindar:2016km}%
  \BibitemOpen
  \bibfield  {author} {\bibinfo {author} {\bibfnamefont {A.~J.}\ \bibnamefont
  {Shalindar}}, \bibinfo {author} {\bibfnamefont {P.~T.}\ \bibnamefont
  {Webster}}, \bibinfo {author} {\bibfnamefont {B.~J.}\ \bibnamefont
  {Wilkens}}, \bibinfo {author} {\bibfnamefont {T.~L.}\ \bibnamefont {Alford}},
  \ and\ \bibinfo {author} {\bibfnamefont {S.~R.}\ \bibnamefont {Johnson}},\
  }\href@noop {} {\bibfield  {journal} {\bibinfo  {journal} {Journal of Applied
  Physics}\ }\textbf {\bibinfo {volume} {120}},\ \bibinfo {pages} {145704}
  (\bibinfo {year} {2016})}\BibitemShut {NoStop}%
\bibitem [{\citenamefont {Sun}\ \emph {et~al.}(2015)\citenamefont {Sun},
  \citenamefont {Ruzsinszky},\ and\ \citenamefont {Perdew}}]{Sun:2015ef}%
  \BibitemOpen
  \bibfield  {author} {\bibinfo {author} {\bibfnamefont {J.}~\bibnamefont
  {Sun}}, \bibinfo {author} {\bibfnamefont {A.}~\bibnamefont {Ruzsinszky}}, \
  and\ \bibinfo {author} {\bibfnamefont {J.~P.}\ \bibnamefont {Perdew}},\
  }\href@noop {} {\bibfield  {journal} {\bibinfo  {journal} {Physical Review
  Letters}\ }\textbf {\bibinfo {volume} {115}},\ \bibinfo {pages} {64}
  (\bibinfo {year} {2015})}\BibitemShut {NoStop}%
\bibitem [{\citenamefont {Heyd}\ \emph {et~al.}(2003)\citenamefont {Heyd},
  \citenamefont {Scuseria},\ and\ \citenamefont {Ernzerhof}}]{Heyd:2003eg}%
  \BibitemOpen
  \bibfield  {author} {\bibinfo {author} {\bibfnamefont {J.}~\bibnamefont
  {Heyd}}, \bibinfo {author} {\bibfnamefont {G.~E.}\ \bibnamefont {Scuseria}},
  \ and\ \bibinfo {author} {\bibfnamefont {M.}~\bibnamefont {Ernzerhof}},\
  }\href@noop {} {\bibfield  {journal} {\bibinfo  {journal} {The Journal of
  Chemical Physics}\ }\textbf {\bibinfo {volume} {118}},\ \bibinfo {pages}
  {8207} (\bibinfo {year} {2003})}\BibitemShut {NoStop}%
\bibitem [{\citenamefont {Cohen}\ \emph {et~al.}(2011)\citenamefont {Cohen},
  \citenamefont {Mori-S{\'a}nchez},\ and\ \citenamefont {Yang}}]{Cohen:2011fm}%
  \BibitemOpen
  \bibfield  {author} {\bibinfo {author} {\bibfnamefont {A.~J.}\ \bibnamefont
  {Cohen}}, \bibinfo {author} {\bibfnamefont {P.}~\bibnamefont
  {Mori-S{\'a}nchez}}, \ and\ \bibinfo {author} {\bibfnamefont
  {W.}~\bibnamefont {Yang}},\ }\href@noop {} {\bibfield  {journal} {\bibinfo
  {journal} {Chemical Reviews}\ }\textbf {\bibinfo {volume} {112}},\ \bibinfo
  {pages} {289} (\bibinfo {year} {2011})}\BibitemShut {NoStop}%
\bibitem [{\citenamefont {Mostofi}\ \emph {et~al.}(2014)\citenamefont
  {Mostofi}, \citenamefont {Yates}, \citenamefont {Pizzi}, \citenamefont {Lee},
  \citenamefont {Souza}, \citenamefont {Vanderbilt},\ and\ \citenamefont
  {Marzari}}]{Mostofi:2014hw}%
  \BibitemOpen
  \bibfield  {author} {\bibinfo {author} {\bibfnamefont {A.~A.}\ \bibnamefont
  {Mostofi}}, \bibinfo {author} {\bibfnamefont {J.~R.}\ \bibnamefont {Yates}},
  \bibinfo {author} {\bibfnamefont {G.}~\bibnamefont {Pizzi}}, \bibinfo
  {author} {\bibfnamefont {Y.-S.}\ \bibnamefont {Lee}}, \bibinfo {author}
  {\bibfnamefont {I.}~\bibnamefont {Souza}}, \bibinfo {author} {\bibfnamefont
  {D.}~\bibnamefont {Vanderbilt}}, \ and\ \bibinfo {author} {\bibfnamefont
  {N.}~\bibnamefont {Marzari}},\ }\href@noop {} {\bibfield  {journal} {\bibinfo
   {journal} {Computer Physics Communications}\ }\textbf {\bibinfo {volume}
  {185}},\ \bibinfo {pages} {2309} (\bibinfo {year} {2014})}\BibitemShut
  {NoStop}%
\bibitem [{\citenamefont {Gmitra}\ and\ \citenamefont
  {Fabian}(2016)}]{Gmitra:2016gf}%
  \BibitemOpen
  \bibfield  {author} {\bibinfo {author} {\bibfnamefont {M.}~\bibnamefont
  {Gmitra}}\ and\ \bibinfo {author} {\bibfnamefont {J.}~\bibnamefont
  {Fabian}},\ }\href@noop {} {\bibfield  {journal} {\bibinfo  {journal}
  {Physical Review B}\ }\textbf {\bibinfo {volume} {94}},\ \bibinfo {pages}
  {1109} (\bibinfo {year} {2016})}\BibitemShut {NoStop}%
\bibitem [{\citenamefont {Kim}\ \emph {et~al.}(2009)\citenamefont {Kim},
  \citenamefont {Hummer},\ and\ \citenamefont {Kresse}}]{Kim:2009dj}%
  \BibitemOpen
  \bibfield  {author} {\bibinfo {author} {\bibfnamefont {Y.-S.}\ \bibnamefont
  {Kim}}, \bibinfo {author} {\bibfnamefont {K.}~\bibnamefont {Hummer}}, \ and\
  \bibinfo {author} {\bibfnamefont {G.}~\bibnamefont {Kresse}},\ }\href@noop {}
  {\bibfield  {journal} {\bibinfo  {journal} {Physical Review B}\ }\textbf
  {\bibinfo {volume} {80}},\ \bibinfo {pages} {035203} (\bibinfo {year}
  {2009})}\BibitemShut {NoStop}%
\bibitem [{\citenamefont {Wu}\ \emph {et~al.}(2018)\citenamefont {Wu},
  \citenamefont {Zhang}, \citenamefont {Song}, \citenamefont {Troyer},\ and\
  \citenamefont {Soluyanov}}]{Wu:2018en}%
  \BibitemOpen
  \bibfield  {author} {\bibinfo {author} {\bibfnamefont {Q.}~\bibnamefont
  {Wu}}, \bibinfo {author} {\bibfnamefont {S.}~\bibnamefont {Zhang}}, \bibinfo
  {author} {\bibfnamefont {H.-F.}\ \bibnamefont {Song}}, \bibinfo {author}
  {\bibfnamefont {M.}~\bibnamefont {Troyer}}, \ and\ \bibinfo {author}
  {\bibfnamefont {T.~G. P. h. p. M. A.~A.}\ \bibnamefont {Soluyanov}},\
  }\href@noop {} {\bibfield  {journal} {\bibinfo  {journal} {Computer Physics
  Communications}\ }\textbf {\bibinfo {volume} {224}},\ \bibinfo {pages} {405}
  (\bibinfo {year} {2018})}\BibitemShut {NoStop}%
\end{thebibliography}%

\end{document}